\documentclass[aps,prd,showpacs,twocolumn,showkeys,preprintnumbers]{revtex4}
\usepackage{graphicx}                              % Used for includegraphics
\usepackage{amsmath,amsfonts}
\graphicspath{{./picturesComp/}}                       % location for color fig.

\usepackage{epsfig}
\usepackage{amssymb}
\usepackage{amsfonts}
\usepackage{float}
\newcommand{\etal}{{\em et al.}\hs{1mm}}

\newcommand{\vs}{\vspace}
\newcommand{\hs}{\hspace}

\newcommand{\bdm}{\begin{displaymath}}
\newcommand{\edm}{\end{displaymath}}
\newcommand{\beq}{\begin{equation}}
\newcommand{\eeq}{\end{equation}}
\newcommand{\bea}{\begin{eqnarray}}
\newcommand{\eea}{\end{eqnarray}}
\newcommand{\bit}{\begin{itemize}}
\newcommand{\eit}{\end{itemize}}
\newcommand{\bc}{\begin{center}}
\newcommand{\ec}{\end{center}}
\newcommand{\re}{\relax{\rm I\kern-.18em R}}

\newcommand{\pol}{\ell} %\wp
\newcommand{\polAbsAvg}{<\fhs{-1.0mm}|\ell|\fhs{-1.0mm}>} %\wp
\newcommand{\polAvgV}{<\fhs{-1.0mm}\ell\fhs{-1.0mm}>_V} %\wp
\newcommand{\fhs}[1]{\mbox{\hs{#1}}}
\newcommand{\ie}{{\it i.e. }}
\newcommand{\x}{\fhs{-0.5mm}}

\begin{document}
\preprint{HU-EP-06/18}

\title{An $SU(2)$ KvBLL caloron gas model and confinement}

\author{P. Gerhold, E.-M. Ilgenfritz, and M. M\"uller-Preussker}
\affiliation{Humboldt-Universit\"at zu Berlin, Institut f\"ur Physik, 
Newtonstr. 15, D-12489 Berlin, Germany}

\date{July 27, 2006}

\begin{abstract}
A semi-classical model is developed to describe pure $SU(2)$
Yang-Mills gluodynamics at finite temperature as a dilute,
non-interacting gas of Kraan-van Baal-Lee-Lu calorons including the case of
non-trivial holonomy.
Temperature dependent parameters of the model (asymptotic caloron
holonomy, caloron density and caloron size distribution) are
discussed from the point of view of lattice observations and
of {\it in-medium} modifications of the one-loop caloron amplitude.
Space-like string tensions running into plateaux at distances
$R \approx 0.8 - 1.3\,\mathrm{fm}$
are obtained and compared to lattice results in order to find more precisely
the average caloron size. Then, the quark-antiquark free energy
as predicted by the model is considered. In the confined phase
a linear rise with the separation can be observed up to
$R \approx 4\, \mathrm{fm}$, whereas it runs into
plateaux above $T_c$. Screening effects in the adjoint potentials
are observed together with an approximate Casimir scaling of the
caloron contribution to the fundamental and adjoint forces.
In Abelian projection, space-like percolation of monopoles is found
in the confined phase only. Thus, taking the non-trivial holonomy
into account, confinement properties of pure $SU(2)$ Yang-Mills
gluodynamics can be described by a semi-classical approach up to
distances one order of magnitude larger than the caloron size.
\end{abstract}

\maketitle

\section{Introduction}
\label{sec:Introduction}

Lattice QCD provides an {\it ab initio} description of fundamental 
features of the hadronic world. It would be useful, however, to have a continuum 
model of the vacuum structure able to reproduce a variety of  lattice results 
and eventually allowing to compute other quantities not accessible in 
Euclidean lattice QCD. 

What we have in mind here is a reference model that could create an ensemble of 
non-Abelian gauge fields, not just various Abelian or center projections thereof, 
analogous to the ansatz by Callan, Dashen and Gross~\cite{Callan1}, who proposed 
to describe the QCD vacuum as a superposition of a dilute set of instantons. 
This approach, widely known as 'dilute gas approximation' (DGA), was refined by 
many authors. Numerical calculations in such instanton based models have been 
performed by sampling the collective parameters of their building 
blocks~\cite{Belavin1}. 
In the case of the 'random instanton liquid model'~(RILM)~\cite{Shuryak1}
the collective parameters were sampled directly (independently), whereas importance 
sampling was used in the so-called 'interacting instanton 
liquid model'~(IILM)~\cite{Shuryak5,Shuryak6}
in order to account for the residual interactions between the pseudoparticles, 
which are induced by temperature and/or the exchange of fermions. 

More recently, superpositions of regular gauge instantons, which are strongly 
interacting due to the slow decay of $A_{\mu}(x)$, have been 
considered for which importance sampling is indispensable~\cite{Negele1}.
In the present paper we wish to extend the RILM in another direction, choosing 
more general pseudoparticles, calorons with generic holonomy, as the basic 
building blocks.

Expanding the Euclidean action around such a solution and using the method of 
collective coordinates~\cite{Gervais1}, it is possible to present the result of 
quantum fluctuations (gauge fixing terms are suppressed) as an expression like
\begin{equation}
Z_1 = e^{-S_E[A^{\rm cl}_\mu]} \int dC \,J(C)\, 
                 \left({\det}^{\prime} M(C)\right)^{-\frac{1}{2}},
\label{eq:oneinstanton}
\end{equation}
which is the single-pseudoparticle contribution to the partition function.
Here ${\det}^{\prime} M(C)$ denotes the non-zero mode determinant of the Hessian 
$M(C)$ of the Euclidean action $S_E[A]$, parametrized by the collective coordinates 
(moduli) $C$ of the classical solution $A^{\rm cl}$. Furthermore, the integrand in 
(\ref{eq:oneinstanton}) includes a Jacobian $J(C)$, which contains the metric $g$ in 
the moduli space in the form $\sqrt{\det g}$ built from the zero modes 
$\frac{\partial A^{\rm cl}}{\partial C}$. As long as interactions are negligible, the 
integrand in (\ref{eq:oneinstanton}) is the probability distribution used for the 
direct sampling of collective parameters for superpositions of solutions of this 
type. The actual density of such lumps must be taken as given by (say, lattice) 
observations. Indeed, a lumpy structure has been observed (and its instanton nature 
taken for granted) in lattice studies using techniques like cooling or 
smoothing~\cite{Ilgenfritz1,Teper1,Hoek1,Polikarpov1,Chu1} or 
smearing~\cite{Hasenfratz1}. 

The example of the instanton liquid shows that the contributions {\it only of small}
instantons are well under perturbative control by the instanton transition amplitude
calculated by 't~Hooft~\cite{tHooft2}. The behavior in the infrared ({\it i. e.} 
the probability of large instantons) is described by other (mainly classical or
Higgs-like) interaction effects dealt with in a mean-field 
fashion~\cite{Ilgenfritz3} or by variational techniques~\cite{Diakonov3}. 
Irrespective of these less determined details, the model defines what we consider 
as a model of semi-classical type, in the sense that the building blocks of the 
model are classical solutions of the Euclidean equations of motion. 

Generically, as the result of mixing of the zero modes of individual instantons, 
the model explains the occurrence of a band of near-zero modes, {\it i.e.} chiral 
symmetry breaking~\cite{Banks1}. The instanton liquid model is successfully 
describing hadronic correlators and details of hadronic 
structure~\cite{Shuryak2,Shuryak3,Shuryak4} 
and is one variant of solving the $U(1)_A$ problem~\cite{tHooft1}. However, without
adding intricate instanton correlations, the instanton liquid model could not 
describe confinement~\cite{Chen1}. This was the motivation to consider strongly 
correlated instantons in the regular gauge~\cite{Negele1}. Our paper is aiming to 
show that the lack of confinement can be overcome, for temperatures in the confining
phase, by the extension of the model from the weakly interacting, singular-gauge 
instanton (caloron) type of classical solutions to more general solutions with 
non-trivial holonomy~\cite{Kraan1,Kraan2,Kraan3,Lee1}.

This extension of the parameter space of the underlying classical solutions can
be motivated by both: analytical considerations~\cite{Diakonov1}, as discussed in 
the following Section, as well as by lattice 
observations~\cite{Shcheredin1,Bruckmann5,Ilgenfritz4,Ilgenfritz5,Gattringer:2002wh,
Gattringer2,Gattringer1,Bruckmann4,Peschka1}.
In Refs.~\cite{Shcheredin1,Ilgenfritz4,Ilgenfritz5} $SU(2)$ 
lattice gauge fields have been analyzed
using smearing techniques and studying their monopole cluster structure.
Below the critical temperature a part of the emerging topological clusters 
could then be characterized to correspond either to non-static calorons or 
to static dyons in the context of Kraan-van Baal-Lee-Lu caloron solutions with 
non-trivial holonomy~\cite{Kraan1,Kraan2,Kraan3,Lee1}. The relative abundance of
topological objects with non-trivial holonomy was furthermore shown to be 
temperature dependent and to decrease above the critical temperature. 
In Refs.~\cite{Gattringer2,Gattringer1} $SU(3)$ Monte-Carlo gauge
fields have been examined by studying the fermionic zero modes. 
Configurations with topological
charge ${Q\x=\x\pm 1}$ probably indicating a 3-dyon structure have been identified. 
The 3-dyon structure
is a direct feature of Kraan-van Baal-Lee-Lu caloron solutions in $SU(3)$ with 
non-trivial holonomy, while 
it can neither be explained as appropriate embeddings of $SU(2)$ instantons 
into $SU(3)$, nor
with calorons with trivial holonomy. A full characterization of $SU(3)$ caloron
solutions by cooling techniques has been given in Ref.~\cite{Peschka1}.
 
This paper is organized as follows: In Section~\ref{sec:Solutions} we will briefly 
describe the new type of classical solutions, calorons of generically non-trivial 
holonomy, to be used in the extension of the instanton liquid model.
In Section~\ref{sec:Superpositions} we will describe the problems encountered in 
the construction of superpositions of calorons with non-trivial holonomy.
In Section~\ref{sec:holonomy_decides} we will demonstrate that, under otherwise 
similar conditions, non-trivial {\it vs.} trivial holonomy determines whether the 
caloron gas confines or not. In Section~\ref{sec:Inputparameters} we will discuss 
realistic input parameters for the model, in particular the analog of the instanton 
size distribution, {\it i. e.} the distribution of the dipole moment of the calorons
in terms of their constituent monopoles. 

Already here a general remark is in order for the orientation of the reader: all 
this could be realized in a continuum model. For practical reasons, however, we 
will discretize the generated gauge field configurations on a suitable grid such 
that the use of lattice techniques becomes possible.
In Section~\ref{sec:Confinementresults} the simulation results for the spatial 
string tensions and the color averaged free energies obtained from our model 
will be presented, before we proceed to Section~\ref{sec:monopoles} where we will 
discuss indirect indicators for confinement observed 
in the monopole structure of the 
generated caloron ensembles (via maximal Abelian gauge and Abelian projection).
In Section~\ref{sec:Outlook} we shall draw conclusions and give an outlook of
what should be done next.

\section{Classical solutions of $SU(2)$ Yang-Mills theory}
\label{sec:Solutions}

The instanton, discovered in 1975~\cite{Belavin1}, is a classical 
solution of the Euclidean Yang-Mills equation of motion at zero temperature with 
localized action density, carrying one unit of topological charge. For any number 
of colors, the instanton is basically an $SU(2)$ object that is parametrized by 
${4\x\cdot\x N_{\rm color}\x=\x8}$ collective coordinates, which are its four-dimensional 
position in space-time,
a size parameter and three parameters describing a global rotation in $SU(2)$ group 
space. For higher $N_{\rm color}$ the additional parameters describe the embedding 
of $SU(2)$ into $SU(N_{\rm color})$. Classical solutions with higher topological 
charge can in principle be constructed by means of the ADHM formalism~\cite{ADHM1},
although they are not analytically available in general. 

\begin{figure*}[htbp]
\centering
\mbox{
\begin{minipage}{0.5\linewidth}
\includegraphics[angle=0,width=1.0\textwidth]{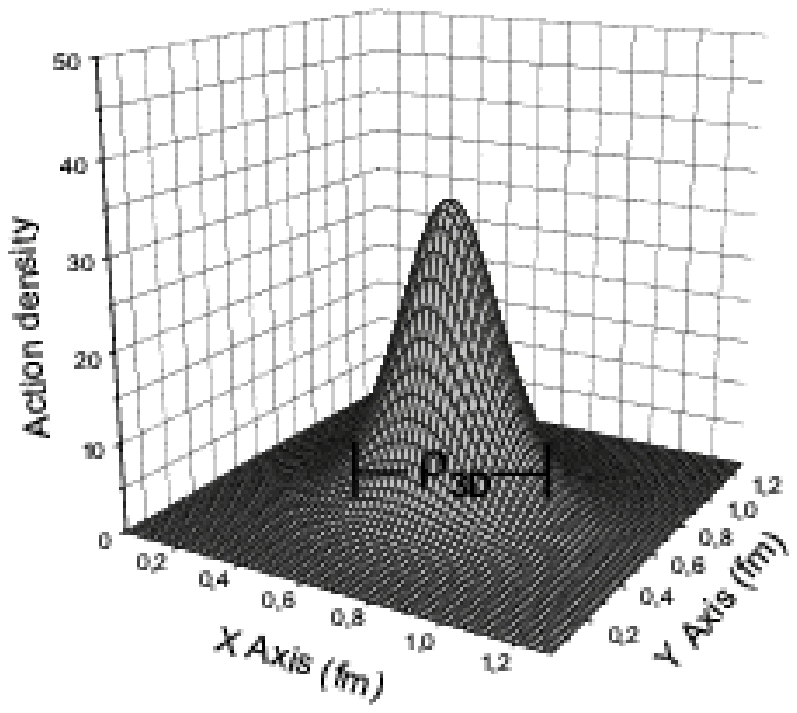}
\end{minipage}
\begin{minipage}{0.50\linewidth}
\includegraphics[angle=0,width=1.0\textwidth]{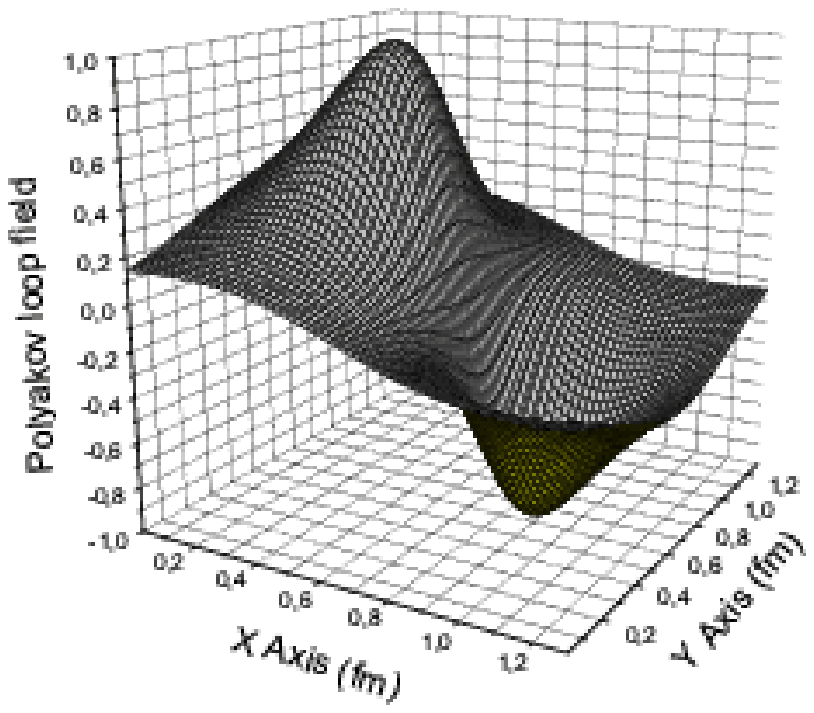}
\end{minipage}
}
\caption[Action density and Polyakov loop distribution of a KvBLL caloron]
{Action density (left) and Polyakov loop distribution (right)
of a KvBLL caloron with maximally non-trivial holonomy ${|\vec\omega|=0.25}$, 
${\rho=0.33}$~fm and ${\beta = 1}$~fm, 
where the action density is given in instanton 
units $S_I \times {\rm fm}^4$. Here, $\rho_{3D}$ denotes the 3D-radius of the 
almost $O(4)$-symmetric
caloron action lump.}
\label{fig:30C}
\end{figure*}

The analog of the instanton in the case of non-zero temperature 
${T\fhs{-0.5mm}=\fhs{-0.5mm}1/\beta\fhs{-0.5mm}\neq\fhs{-0.5mm} 0}$ 
was discovered soon after the $T=0$ instanton by Harrington and Shepard in 
1978~\cite{Harrington1} and was accepted as an appropriate semi-classical 
background at finite temperature. It will be referred to as ``HS caloron'' 
throughout the present paper. 

Twenty years later, Kraan and van Baal~\cite{Kraan1,Kraan2} and 
Lee and Lu~\cite{Lee1} 
extended the parameter space of the HS caloron by an additional parameter, 
the asymptotic holonomy 
${\cal P}_{\infty}$ which is defined as
\begin{equation}
{\cal P}_{\infty} = e^{2\pi i \vec \omega \vec \tau} = 
\lim\limits_{|\vec x| \rightarrow \infty} P(\vec x) \,,
\label{eq:holonomy}
\end{equation}
where $\vec\tau$ denotes the vector of Pauli matrices and 
\begin{equation}
P(\vec x) = \hat P\, \exp\left( i \int\limits_0^\beta A_4(\vec x,t) dt\right) \in SU(2)\,,  
\label{eq:Polyakov}
\end{equation}
where path ordering is implied by the $\hat P$ symbol.
The generalized solution, which we will call KvBLL caloron, is reduced to the HS 
caloron for trivial asymptotic holonomy, taking values in the center of the gauge 
group, 
${\cal P}_{\infty} \in Z(2)$. In the case of general $N_{\rm color}$, this solution 
can also be constructed. It is not a simple embedding of an $SU(2)$ 
solution into the bigger group. The present paper, however, 
is restricted to the case of $N_{\rm color}=2$.

Soon after the discovery of the respective solutions, instanton ensembles as well 
as HS caloron ensembles with trivial holonomy have been used to model the QCD 
vacuum at T=0 or at finite temperature, respectively. Concerning non-trivial 
holonomy, the ruling opinion was since the beginning of the 1980's that such 
classical solutions cannot play a significant role in the QCD partition function. 
This was due to an argument given by Gross, Pisarski and Yaffe~\cite{Gross1} that 
gauge fields with non-trivial holonomy are exponentially suppressed by the
perturbative free energy~\cite{Weiss1}. Hence, their
potential relevance for quark confinement has not been explored so far. However,
one has to keep in mind that the argument does not hold for caloron ensembles
with a finite density of calorons in space.

Very recently, Diakonov \etal have calculated the holonomy dependence of the 
free energy of a non-interacting KvBLL caloron gas as a consequence of the 
caloron quantum weight~\cite{Diakonov1}. For sufficiently high temperatures, 
trivial holonomy was shown to minimize the free energy and seems therefore dominant 
in the deconfined phase. However, it becomes unstable below a certain 
temperature. Thus, one may conjecture that KvBLL calorons with non-trivial holonomy 
may become the relevant degrees of freedom for lower temperatures. With this new 
argument in mind and with the KvBLL caloron solution at hand it seems  a natural 
step to study KvBLL caloron gas models.

The KvBLL caloron with arbitrary asymptotic holonomy is a self-dual gauge field, 
{\it i.e.} ${F_{\mu\nu}\fhs{-0.5mm}=\fhs{-0.5mm}\tilde F_{\mu\nu}}$ with 
${\tilde F_{\mu\nu}\fhs{-0.5mm}=\fhs{-0.5mm}\frac{1}{2} 
\epsilon_{\mu\nu\alpha\beta} F_{\alpha\beta}}$,
solving the Euclidean equation of motion at finite temperature. It was first 
constructed in the so-called algebraic gauge by means of the ADHM 
formalism~\cite{ADHM1} by Kraan and van Baal~\cite{Kraan1,Kraan2} and by Lee and 
Lu~\cite{Lee1} as a self-dual gauge field in flat $\re^4$ with the 
periodicity condition 
$A^{alg}_\mu(x+\beta \hat e_t)= {\cal P}_{\infty} A^{alg}_\mu(x) 
{\cal P}_{\infty}^{\dagger}$, 
where $\hat e_t$ denotes the unit vector along the time direction. The resulting 
vector potential $A_\mu^{alg}$ defined on $\re^4$ can then be transformed into a 
self-dual and time-periodic gauge field $A_\mu^{per}$ defined on $S^1\times \re^3$, 
by a gauge transformation $\Omega(x)$ non-periodic in time  
\begin{eqnarray}
A_\mu^{per}(\vec x, t\, \mbox{mod}\, \beta) &=& 
\Omega(x) A^{alg}_\mu(\vec x,t)\Omega^\dagger(x) \nonumber\\
&-& i\,\Omega(x)\partial_\mu\Omega^\dagger(x) \; ,
\label{eq:eq30F}
\end{eqnarray}
with $x=(\vec x, t)$ and
\begin{equation}
\Omega(x) = e^{- 2\pi i\vec\tau \vec\omega t/\beta} \; . 
\label{eq:Omega}
\end{equation}
This field is carrying one unit of topological charge and the asymptotic holonomy 
${\cal P}_{\infty}$. 
It is the KvBLL caloron in the periodic gauge with arbitrary asymptotic holonomy 
${{\cal P}_{\infty} = \exp(2\pi i \vec \omega\vec\tau) \in SU(2)}$,
and one finds that the KvBLL caloron is also described by 8 collective coordinates 
in addition to its holonomy parameter $\vec \omega$. However, the physical 
interpretation of these collective coordinates is different from that of the 
instanton parameters, as will be discussed in the following.

One can imagine the KvBLL caloron in the algebraic gauge as an infinite chain of 
equally separated, identical instantons in flat $\re^4$, aligned along the time 
direction with each subsequent instanton rotated by the 
holonomy ${\cal P}_{\infty}$ relative to the preceding one - 
as long as the size $\rho$ of the instantons is small compared to the inverse 
temperature $\beta$. In that case it is appealing to parametrize the caloron in 
terms of the collective coordinates of the instanton.
For $\rho\ll\beta$ the corresponding caloron consists
of one lump of action, which is approximately $O(4)$-rotationally symmetric in 
space-time, and its radius can be described by the parameter $\rho$. 
Fig.~\ref{fig:30C} shows the action density and the field of the Polyakov loop 
$\pol(x)$, where 
\begin{equation}
\pol(\vec x) = \frac{1}{N_{\rm color}} Tr\,P(\vec x) \; . 
\label{eq:Polyakovfield}
\end{equation}
This example represents a caloron with maximally 
non-trivial holonomy and ${\rho\fhs{-0.5mm}\ll\fhs{-0.5mm}\beta}$. 
The Polyakov loop goes to zero at spatial infinity according to the maximally 
non-trivial holonomy.

\begin{figure*}[htbp]
\centering
\mbox{
\begin{minipage}{0.5\linewidth}
\includegraphics[angle=0,width=1.0\textwidth]{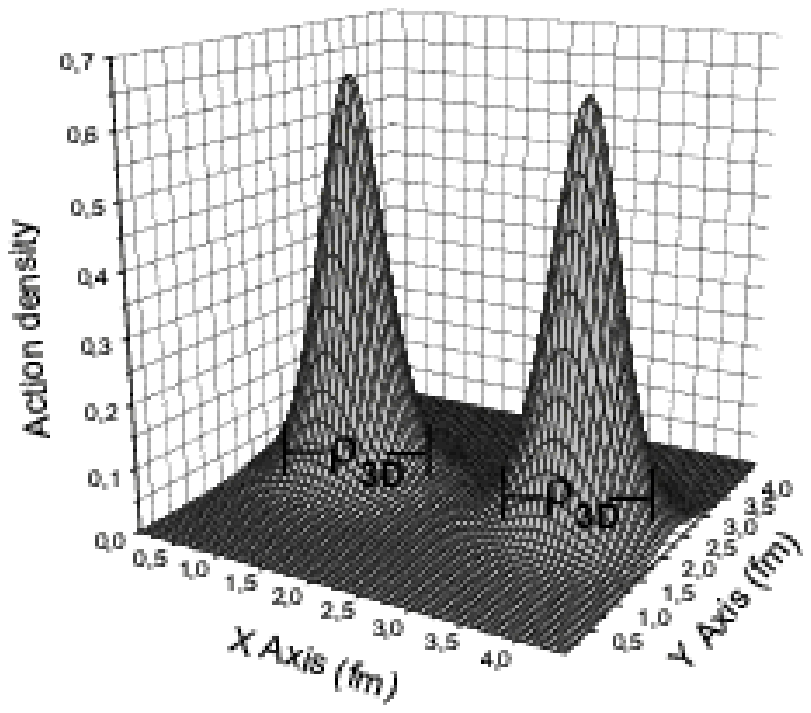}
\end{minipage}
\begin{minipage}{0.50\linewidth}
\includegraphics[angle=0,width=1.0\textwidth]{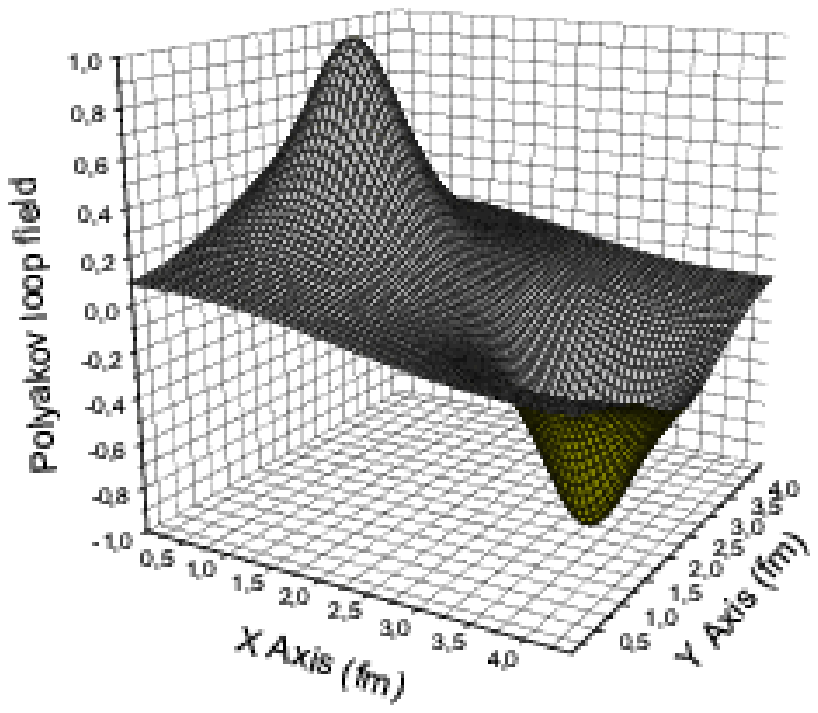}
\end{minipage}
}
\caption{Action density (left) and Polyakov loop distribution (right) 
of a dissociated KvBLL caloron with maximally non-trivial holonomy 
${|\vec\omega|=0.25}$, ${\rho=1}$~fm and ${\beta = 1}$~fm, where the action density 
is given in instanton units $S_I \times {\rm fm}^4$. For growing $\rho$ two constituents 
emerge from the spherical action lump shown in Fig.~\ref{fig:30C}. 
Here, $\rho_{3D}$ denotes the radius of the almost $O(3)$-symmetric 
monopole action 
lumps in 3 dimensions.}
\label{fig:30D}
\end{figure*}

In the opposite case, ${\rho\fhs{-0.5mm}\gg\fhs{-0.5mm}\beta}$, the caloron 
is dissociated into two constituents, as shown in the action density plot in 
Fig.~\ref{fig:30D}. For ${\rho\fhs{-0.5mm}\rightarrow\fhs{-0.5mm}\infty}$ these 
constituents become static in time in an appropriate gauge, breaking down the prior 
$O(4)$ symmetry to an $O(3)$-rotational symmetry in 3-dimensional position space, 
and the corresponding 3-dimensional radii $\rho_{3D}^{(1,2)}$ of the two 
constituents converge to constants approximately given by 
${\rho_{3D}^{(1)}\approx \beta/\omega}$ and 
${\rho_{3D}^{(2)}\approx\beta/\bar\omega}$, respectively, where 
${\omega=|\vec \omega|\in [0;1/2]}$ and ${\bar\omega = 1/2-\omega}$. The 
constituents can then be identified with BPS monopoles~\cite{Prasad1} each carrying 
the fraction $2\omega$ or $2\bar\omega$, respectively, of the total action and 
topological charge. The distance between the monopole positions 
$\vec z_1$, $\vec z_2$ is given by 
${d\fhs{-0.5mm} =
\fhs{-0.5mm} |\vec z_1 - \vec z_2|\fhs{-0.5mm} =\fhs{-0.5mm} \pi\rho^2/\beta}$. 
Due to the emergence of these two monopoles 
at larger $\rho$ it seems more natural to use their 
positions $\vec z_1$ and $\vec z_2$ for the parametrization of the caloron instead 
of the instanton parameters involved in the construction scheme. The corresponding
vector potentials in the algebraic and the periodic gauge are then analytically
given by~\cite{Kraan1,Kraan2}
\begin{eqnarray}
A^{alg}_\mu(x)\x\x &=&\x\x \frac{\phi}{2}
Re\left[(\bar\eta^1_{\mu\nu}-i\bar\eta^2_{\mu\nu})(\tau_1+i\tau_2)\partial_\nu \chi   
\right] \nonumber\\
\x\x&+&\x\x \frac{\tau_3}{2} \bar\eta_{\mu\nu}^3 \partial_\nu \ln\phi , 
\label{eq:eq30I}
\end{eqnarray}
\vs{-5mm}
\begin{eqnarray}
A^{per}_\mu(x)\x\x &=&\x\x \frac{\phi}{2}
Re[(\bar\eta^1_{\mu\nu}-i\bar\eta^2_{\mu\nu})(\tau_1+i\tau_2)\nonumber\\
\x\x&\times&\x\x(\partial_\nu + 4\pi i \omega\delta_{\nu,4})\tilde\chi  ]  \nonumber \\
\x\x&+&\x\x \frac{\tau_3}{2} 
\bar\eta_{\mu\nu}^3 \partial_\nu \ln\phi + \delta_{\mu,4}2\pi \omega \tau_3 \; , 
\label{eq:eq30H}
\end{eqnarray}
with the functions $\psi(x)$, $\hat \psi(x)$, $\phi(x)$, $\chi(x)$, 
$\tilde\chi(x)$ defined according to
\begin{eqnarray}
\psi(x)\x\x &=&\x\x \frac{r^2+s^2+\pi^2\rho^4}{2rs}
\sinh(4\pi r\bar\omega)\sinh(4\pi s \omega)\nonumber\\
\x\x&+&\x\x \cosh(4\pi r\bar\omega) \cosh(4\pi s\omega) -\cos(2\pi t) \nonumber\\
\x\x&+&\x\x \pi\rho^2 s^{-1}\sinh(4\pi s \omega)\cosh(4\pi r\bar\omega) \nonumber\\
\x\x&+&\x\x \pi\rho^2 r^{-1}\sinh(4\pi r \bar\omega)\cosh(4\pi s\omega) \,,   
\label{eq:psi}
\end{eqnarray}
\vs{-5mm}
\begin{eqnarray}
\hat \psi(x)\x\x &=&\x\x \frac{r^2+s^2-\pi^2\rho^4}{2rs}
\sinh(4\pi r \bar\omega)\sinh(4\pi s\omega)   \nonumber\\
\x\x&+&\x\x \cosh(4\pi r\bar\omega)\cosh(4\pi s\omega) - \cos(2\pi t), 
\label{eq:psihat}
\end{eqnarray}
\vs{-5mm}
\begin{eqnarray}
\chi(x)\x\x &=&\x\x e^{4\pi it\omega}\frac{\pi\rho^2}{\psi} 
s^{-1}\sinh(4\pi s \omega)e^{-2\pi i t} \nonumber\\
\x\x&+&\x\x   e^{4\pi it\omega}\frac{\pi\rho^2}{\psi}   r^{-1}\sinh(4\pi r \bar\omega) \,,
\label{eq:chi}
\end{eqnarray}
\vs{-5mm}
\begin{eqnarray}
\tilde\chi(x)\x\x &=&\x\x e^{-4\pi i t \omega}\chi \,,
\label{eq:chitilde}
\end{eqnarray}
\vs{-5mm}
\begin{eqnarray}
\phi(x)\x\x &=&\x\x \frac{\psi}{\hat \psi} \, .
\label{eq:eq30E}
\end{eqnarray}
Here ${\vec r\x=\x\vec x\x-\x\vec z_1}$, ${\vec s\x=\x\vec x\x-\x\vec z_2}$, 
${r\x=\x|\vec r|}$, ${s\x=\x|\vec s|}$, and ${\bar\eta^a_{\mu\nu}}$ denotes 
the 't Hooft symbol~\cite{tHooft2}.
For convenience, the asymptotic holonomy parameter was set equal to 
${\vec\omega\vec\tau\x=\x\omega\tau_3}$, the inverse temperature was chosen 
equal to unity, ${\beta\x=\x1}$, and the constituents were assumed to be 
separated along the $z$-axis, \ie ${\vec z_1,\vec z_2 \parallel \hat e_z}$, with 
their $t$-coordinates set equal to zero. The general gauge field 
with arbitrary monopole positions, arbitrary temperature and holonomy 
${{\cal P}_{\infty}\fhs{-0.5mm}=\fhs{-0.5mm}\exp(2\pi i \vec\omega\vec\tau)}$ 
can be obtained by applying appropriate global transformations on the 
vector potential. 
These are a spatial rotation, a translation in space-time, a global gauge 
rotation $\Omega$ in color space, accomplishing
${\Omega\omega\tau_3\Omega^\dagger\fhs{-0.5mm}=\fhs{-0.5mm}\vec\omega\vec\tau}$, 
and an adequate rescaling. Hence, the KvBLL caloron for a given temperature and 
fixed holonomy can be parametrized by the four-dimensional position of its center, 
the size parameter $\rho$ determining the monopole separation, two angles 
describing their spatial rotation and one parameter for an residual $U(1)$ gauge 
rotation around the axis $\vec\omega\vec\tau$ in color space. This is the 
parametrization that will be applied in the present work. 
Three further remarks shall be given here.

{\em (i)} In the case of non-trivial asymptotic holonomy ${\cal P}_{\infty}$ 
the caloron
gauge field does not vanish at spatial infinity. In the periodic gauge the $A_4$ 
component of the vector potential converges to a constant at spatial infinity 
whereas the remaining three components go to zero, 
\begin{equation}
\lim\limits_{|\vec x|\rightarrow \infty}A^{per}_4(x) = 2\pi \vec\omega 
\vec\tau, \mbox{\hs{1mm}} \lim\limits_{|\vec x|\rightarrow \infty} 
A^{per}_{1,2,3}(x) = 0 \; .
\label{eq:eq30G}
\end{equation}
The non-vanishing vector potential will become an issue when one tries to 
superpose KvBLL calorons.

{\em (ii)} In the case ${\omega\fhs{-0.5mm}\rightarrow\fhs{-0.5mm} 0}$ or 
${\bar\omega\fhs{-0.5mm}\rightarrow\fhs{-0.5mm} 0}$ the 
asymptotic holonomy becomes trivial, one constituent vanishes and the 
remaining one becomes a HS caloron. 

{\em (iii)} Finally, the anticaloron, which is the anti-self-dual analogue of the 
caloron, carrying negative topological charge, can be obtained from the 
caloron gauge field by 
\begin{eqnarray}
A^{anti}_i (\vec x, t) &=& -A_i(-\vec x,t) \; , \quad i=1,2,3 , \nonumber\\
A^{anti}_4 (\vec x, t) &=& A_4(-\vec x,t) \; .
\label{eq:dualtoantidual}
\end{eqnarray}

\section{Superpositions of calorons}
\label{sec:Superpositions}

There has been remarkable progress in the construction of KvBLL calorons with 
higher topological charge~\cite{Bruckmann1,Bruckmann2}. Analytical expressions 
for an arbitrary number of calorons with their monopoles placed along one axis 
have been derived. Analytical parametrizations for the whole 
parameter space of classical solutions with arbitrary topological charge,
however, are not available in general. Furthermore, configurations of mixed 
self-dual/anti-self-dual character are not known as solutions of the equations 
of motion, just those configurations which
are of overwhelming importance. Therefore, the model introduced in this paper 
will be based on approximate classical gauge field configurations, constructed 
out of single caloron and anticaloron solutions by some kind of superposition 
scheme. In the case of the ``random instanton liquid'' this scheme consisted 
simply of adding $N$ single (anti)instanton gauge fields (``sum-ansatz'')
\beq
A_\mu(x) = \sum\limits_i A_\mu^{(i)}(x), \quad i=1,...,N ,
\label{eq:xxx1}
\eeq
where the $A_\mu^{(i)}(x)$ were chosen in the singular gauge, in which the 
vector potential outside the instanton core drops to zero most quickly. 
Due to the non-linearity of $F_{\mu\nu}$ in terms of the vector potential, 
this ansatz leads to deviations from exact (anti-)self-duality. However, 
since the non-linearity arises from the commutator term $[A_\mu, A_\nu]$, 
the superposition approximately describes a classical solution as long as 
at every space-time position $x$ the sum of the gauge fields is dominated
by the contribution $A_\mu^{(i)}(x)$ of one single caloron, or if all vector 
potentials are almost Abelian. 

For instantons, the simple sum-ansatz yields good approximations of 
multi-instanton configurations, unless the separation between the instantons 
becomes smaller than the size $\rho$ of their action lumps. The quantitative 
criterion ${\rho_1^2\rho_2^2/|x_1-x_2|^4\fhs{-0.5mm}<\fhs{-0.5mm}1/200}$ for 
the applicability of the sum-ansatz for the superposition of two instantons 
with sizes $\rho_1,$ and $\rho_2$ and positions $x_1$ and $x_2$ was established 
in~\cite{Ilgenfritz3}.

When dealing with calorons with non-trivial holonomy this criterion no 
longer holds because the KvBLL caloron gauge field does not vanish even far 
away from the monopole locations. At first, the potential $A_4$ does not vanish 
at spatial infinity in the case of non-trivial holonomy, inducing a correlation 
between calorons even at infinite separations. Secondly, there is a string 
(``Dirac string'') of strong vector potential between the monopoles of a caloron 
with non-trivial holonomy. Despite its strength, this string is almost free of 
action, since its vector potential is ``fine-tuned''. However, there can be 
strong interactions between Dirac strings and further calorons. Therefore, more care 
is required when one is going to apply the simple sum-ansatz to KvBLL calorons 
with non-trivial holonomy, in order to avoid that the overlapping vector potentials 
of two calorons spoil the exact (anti-)self-duality of the single solutions.

The first problem can be overcome quite easily by choosing the algebraic gauge 
for all the calorons before adding their potentials. In this gauge the potential 
$A_4^{alg}$ vanishes at spatial infinity and the sum-ansatz does no longer yield 
interactions between infinitely far separated calorons. On the other hand, the 
vector potentials are then no longer periodic. Periodicity has to be restored 
by applying another gauge transformation to the sum of the vector potentials.
This leads to the modified sum-ansatz 
\bea
A^{per}_\mu(x) &=& e^{-2\pi i t\vec{\omega}\vec{\tau}} \cdot 
\sum_i A^{(i),alg}_\mu(x) \cdot e^{+2\pi i t\vec{\omega}\vec{\tau}} \nonumber\\
&+& 2\pi \vec{\omega}\vec{\tau}\cdot \delta_{\mu,4} .
\label{eq:xxx3}
\eea
However, this superposition scheme is only valid under the restriction that 
only calorons and anticalorons with identical holonomy be superposed because the 
gauge transformation necessary to bring the caloron vector potential from the 
algebraic to the periodic gauge depends on the holonomy according to 
(\ref{eq:eq30F}). It would not be possible to restore the periodicity of a 
configuration of calorons in the algebraic gauge with a global gauge 
transformation, if the configuration were consisting of calorons with 
different holonomies.

The second problem, involving the Dirac strings, will be addressed in an upcoming 
paper and is not discussed here. Although the construction, helpful in this case, 
is of principal interest, it will turn out in the following Section that the 
calorons considered in this model seem to be only weakly dissociated in realistic 
physical situations. They are mostly merged together into joined action lumps. 
This implies that interactions between Dirac strings and other calorons hardly occur. 

As discussed before for the case of instantons, superposing calorons according 
to the sum-ansatz (\ref{eq:xxx3}) becomes unreasonable if the separations between 
different calorons or their constituents, respectively,  become too small. In 
such cases an improved superposition scheme is needed. A good candidate for such 
an improved scheme will also be discussed in the upcoming paper. This improved
superposition procedure can be derived exploiting the ADHM-formalism and restores 
Shuryak's {\it ratio-ansatz}~\cite{Shuryak8} when applied to instantons. However, 
for the caloron model, introduced in the current work, it will turn out that 
the caloron liquid is sufficiently dilute under realistic conditions to 
{\it justify} the sum-ansatz. A measure for the quality of a caloron superposition 
is given by the ``action surplus'' factor 
\beq
\gamma = \frac{S[\sum A_\mu^{(i)}]}{\sum S[A_\mu^{(i)}]}
\label{eq:DefActionSurplus}
\eeq
which becomes unity for exactly (anti-)self-dual configurations, whereas $\gamma >1$, 
if the classical equation of motion is piecewise violated.

So far, the discussion was referring completely to the continuum. In order to 
evaluate a certain set of desired observables numerically, it is convenient 
to introduce a grid (called ``lattice'' in the following). In the later 
calculations, the gauge field $A_\mu(x)$ 
of the caloron ensemble, which is analytically given by the sum-ansatz, 
will be represented by a set of link variables according to 
\beq
U_{x,\mu} = \hat P\,\exp\left( i\int\limits_x^{x+a\hat\mu} 
A_\mu(y) dy_\mu\right) \; .
\label{eq:eq2Y}
\eeq
\begin{figure*}[htb]
\centering
\begin{tabular}{cc}
\hs{6mm}(a) & \hs{6mm}(b)\\
\includegraphics[angle=0,width=0.48\textwidth]{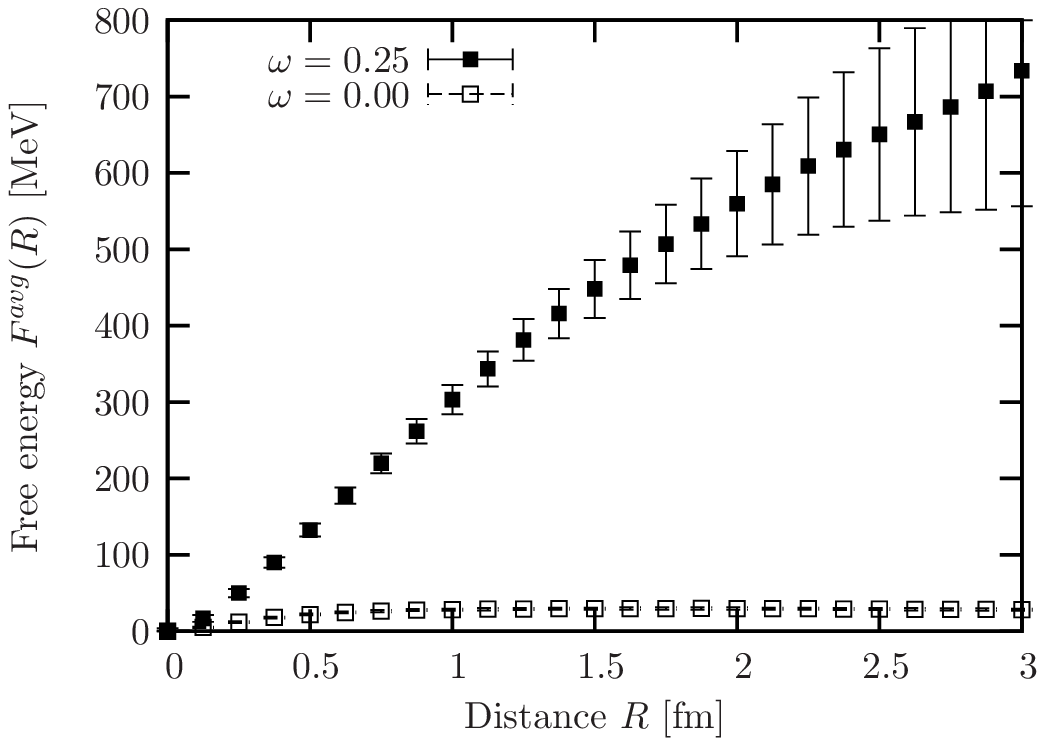} &
\includegraphics[angle=0,width=0.48\textwidth]{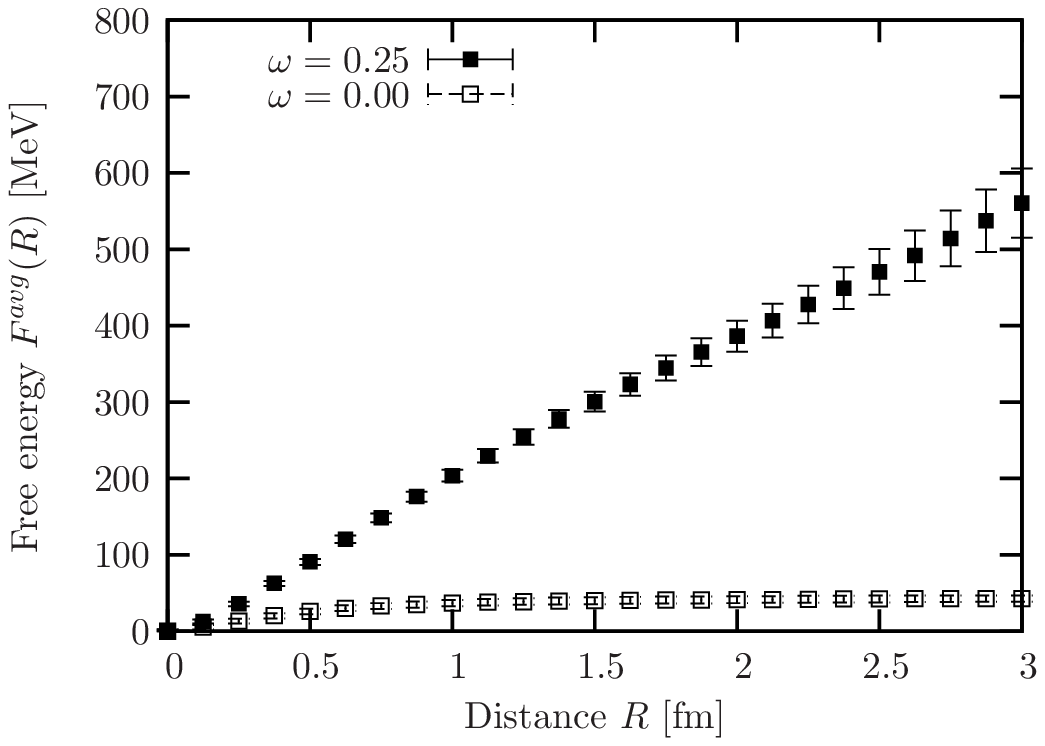}\\
\end{tabular}
\caption{Heavy quark-antiquark free energy $F^{avg}(R)$ derived from Polyakov loop 
correlators measured in caloron ensembles 
($n=1\,\mathrm{fm}^{-4},\, \beta=1\, \mathrm{fm}$) with 
trivial ($\omega=0$) and maximally non-trivial holonomy 
($\omega=0.25$). 
(a) The size parameter is fixed to $\rho=0.33\,\mathrm{fm}$. 
(b) The size parameter is sampled according to a distribution with finite 
width.}
\label{fig:TrivVsNonTriv1}
\end{figure*}
The path ordering operator $\hat P$ will approximately be accounted for by 
dividing each path of integration into $N$ sub-intervals with their lengths 
${a/N\fhs{-0.5mm}\ll\fhs{-0.5mm}\bar\rho}$ being small compared to the average 
caloron size, such that the vector potential along these sub-intervals can be 
assumed to be constant. 
In the numerical calculations the convergence of the link variables was 
ensured by adjusting the number of sub-intervals dynamically, corresponding 
to the variability of the gauge field, and checked by comparing the resulting 
parallel transporter for different values of $N$. 

\section{Caloron ensembles with trivial versus non-trivial holonomy}
\label{sec:holonomy_decides}

Semi-classical studies based on calorons with trivial holonomy, 
{\it i.e.} HS calorons, have already extensively been undertaken
in the past~\cite{Gross1}. Apart from Diakonov's new argument~\cite{Diakonov1}, 
opening the way to consider calorons with non-trivial holonomy, {\it i.e.} 
KvBLL calorons, one may wonder what the most interesting physical objectives are 
to begin simulations of KvBLL caloron ensembles. A first striking demonstration 
what the consequences of expanding the caloron parameter space would be, 
can be given when we compare the heavy quark-antiquark free energy, extracted 
from the Polyakov-loop correlator in a random ensemble of HS calorons, with the same
quantity obtained from random ensembles of KvBLL calorons with maximally non-trivial 
holonomy. 

These ensembles have been created on an open $32^3\times 8$ lattice, embedded in a
bigger continuum of the same temporal periodicity that is randomly filled with 
calorons of a certain density. The links have been generated as described in
the previous Section. The inverse temperature has been set equal to
${\beta\fhs{-0.5mm}=\fhs{-0.5mm}1}$~fm.
Postponing considerations about the realistic choice of the model parameters
to Section \ref{sec:Inputparameters} we have here assumed a caloron density
of ${n\fhs{-0.5mm}=\fhs{-0.5mm}1~{\rm fm}^{-4}}$ and a caloron size parameter
fixed to ${\rho\fhs{-0.5mm}=\fhs{-0.5mm}0.33}$~fm.
The color-averaged heavy quark-antiquark excess free energy $F^{avg}(R)$
can be calculated for the generated configurations by means of the Polyakov
loop correlator according to
\beq
F^{avg}(R) = - \frac{1}{\beta} \ln \left( << \pol(\vec x) \cdot 
 \pol(\vec y) >_V>_C \right) \; ,
\label{eq:eq2ZZS}
\eeq
where ${<\fhs{-1.0mm}\cdot\fhs{-1.0mm}>_C}$ denotes the average over all 
configurations and ${<\fhs{-1.0mm}\cdot\fhs{-1.0mm}>_V}$
the volume average over all pairs of lattice sites $x,y$ with 
${R\fhs{-0.5mm}=\fhs{-0.5mm}|\vec x - \vec y|}$ fixed. 

Fig.~\ref{fig:TrivVsNonTriv1}a shows $F^{avg}$ for trivial 
(${\omega\fhs{-0.5mm}=\fhs{-0.5mm}0}$) 
and maximally non-trivial holonomy (${\omega\fhs{-0.5mm}=\fhs{-0.5mm}0.25}$), 
while all other parameters are kept fixed. 
The striking observation is that the excess free energy of the pair 
rises monotonously with the distance $R$ in the latter case, whereas it runs 
into a plateau, corresponding to the deconfinement of quarks, for the HS 
caloron ensemble in the first case. The effect is directly caused by the
holonomy. This becomes obvious, if one considers the pair free energy 
$F^{avg}$ for asymptotically large distances $R$. 
Then the covariance ${\mathrm{cov}_V(\pol(\vec x),\pol(\vec y))}$ of $\pol(\vec x)$ and 
$\pol(\vec y)$, which is defined as 
${<\fhs{-1.3mm}[<\fhs{-1.5mm}\pol(\vec x)\fhs{-1.5mm}>_V\fhs{-1.5mm}
-\pol(\vec x)]\cdot [<\fhs{-1.0mm}\pol(\vec y)\fhs{-1.0mm}>_V
-\pol(\vec y)]\fhs{-1.5mm}>_V}$, goes to zero and one obtains
\bea
F^{avg}(R) \fhs{-1.0mm}= \fhs{-2.0mm} & - \frac{1}{\beta}&
\ln \left( < \fhs{-1.0mm}\mathrm{cov}_V(\pol(\vec x), \pol(\vec y)) +  
<\fhs{-1.5mm}\pol\fhs{-1.5mm}>_V^2 \fhs{-1.5mm}>_C\right) \nonumber \\
\stackrel{R\rightarrow\infty}{\longrightarrow} \fhs{-1.5mm} & - \frac{1}{\beta}& 
\ln \left( <<\pol>_V^2 >_C \right) \; .
\label{eq:eq2ZZS2}
\eea
While $<\fhs{-1mm}\pol\fhs{-1mm}>_V^2 = 0$ for maximally non-trivial holonomy, 
leading to an infinitely rising quark-antiquark potential,
the average Polyakov loop ${<\fhs{-1mm}\pol\fhs{-1mm}>_V^2}$ is non-zero for 
every configuration with ${\omega\fhs{-0.5mm}\neq\fhs{-0.5mm}0.25}$ due to the
restriction induced by the superposition scheme, hence forcing the potential to 
run into a plateau.

The missing {\it strictly linear} rise of the free energy on the quark separation
even for $\omega=0.25$, which would have been expected at large distances $R$, 
is an unsatisfactory feature of this special example with fixed caloron size. 
It can be ameliorated by sampling the caloron size parameters $\rho$ 
according to a distribution $D(\rho)$ with a finite width. This is demonstrated 
in Fig.~\ref{fig:TrivVsNonTriv1}b, where a $\rho$-distribution 
${D(\rho)\fhs{-0.5mm}\propto\fhs{-0.5mm}\rho^{7/3} \exp(-c\rho^2)}$ has been used, 
which will be physically motivated later. Here, the constant $c$ is determined 
by fixing the average size parameter ${\bar\rho\fhs{-0.5mm}=\fhs{-0.5mm}0.33}$~fm. 
We will return to the question of a realistic $\rho$-distribution in 
Section~\ref{subsec:SizeDis}. 
In conclusion of this Section we can state that for caloron gases with temperature,
density and caloron size in the right ballpark a reasonable confining potential 
becomes simply a matter of maximally non-trivial holonomy.

\section{The parameters of the caloron gas model}
\label{sec:Inputparameters}

In the following, for a realistic $SU(2)$ KvBLL multi-caloron gas model 
the input parameters will be specified to describe pure gluodynamics.
The model is completely defined by describing how the parameters 
of each single caloron are sampled. The 4-dimensional center position, 
the spatial orientation of the two monopoles, 
a global $U(1)$ rotation around the axis $\vec\omega\vec\tau$ in color space 
will be sampled completely randomly, and the ``size'' parameter $\rho$ will be 
sampled according to a suitable $\rho$-distribution $D(\rho,T)$.
The average number of calorons placed on the lattice is determined by the 
caloron density $n(T)$ and the physical four-dimensional volume of the lattice,
one extension of which is given by the time periodicity length 
${\beta\fhs{-0.5mm}=\fhs{-0.5mm}1/T}$. 
Finally, the holonomy parameter $\vec\omega(T)$, which is assumed to be a function
of the temperature, determines the type of  
superposed solutions. According to the selected superposition scheme introduced
in Section \ref{sec:Superpositions} it has to be the same 
for each superposed caloron in the same configuration. 
We put ${\vec\omega(T)\fhs{-0.5mm}\equiv\fhs{-0.5mm}\omega(T)\hat e_3}$ 
without loss of generality. We are aware that this setting breaks the global
$Z(2)$ invariance. It could be easily restored by randomly selecting values
$\omega(T)$ or $\bar\omega(T)= 0.5 - \omega(T)$. 

Expectation values of observables are then obtained from averaging over the 
constructed configurations without additional weighing, since the classical 
action is proportional to the caloron density (apart from local violations 
of the equation of motion) which is the same for all configurations. Quantum 
fluctuations are approximately accounted for in the sampling process 
by an adequate choice of the caloron $\rho$-distribution $D(\rho,T)$.

The physical scale selected by QCD enters the calculation through the 
dimensionful parameters of the model, which are the caloron density $n(T)$
and the $\rho$-distribution $D(\rho,T)$ (both in their dependence on temperature).
Finally, in the deconfined phase, the temperature dependence of 
the holonomy parameter $\omega(T)$ might be important.
These model parameters should be chosen consistently with lattice observations.
This will be described in the following and gives the opportunity to discuss
what is known about these quantities. We should keep in mind that
the confining property is independent of the detailed choice of parameters,
as long as maximally non-trivial holonomy is realized in the confined phase.

\subsection {The holonomy in the confined and deconfined phase}
\label{subsec:DetOfHol}
\vs{1mm}

The KvBLL caloron offers the option to set the asymptotic 
holonomy ${\cal P}_{\infty}$
to an arbitrary value in $SU(2)$ through its $\vec\omega$ parameter. In a 
sufficiently dilute multi-caloron gas the single caloron holonomy parameter
$\vec\omega$ determines the volume-averaged Polyakov loop which is (approximately) 
selfconsistently determining again the holonomy parameter through  
${\polAvgV\fhs{-1.5mm}=\fhs{-0.5mm} cos(2\pi |\vec \omega|)}$. At 
higher densities this simple relation no longer holds, and the average Polyakov loop
is stronger influenced by the internal Polyakov loop profile of the solutions.
It becomes a function of the holonomy parameter $\vec \omega$, the caloron 
density $n$, and the $\rho$-distribution. 
The procedure in the present work is to adopt the holonomy parameter 
and its dependence on temperature directly from lattice results for the 
renormalized Polyakov loop.

Obviously, the average Polyakov loop on the lattice is a bare quantity 
suffering from ultraviolet divergences, since the Polyakov loop is proportional 
to the propagator of an infinitely massive test quark moving along the time 
direction. The additive mass shift for the test quark resulting from the lattice 
regularization affects the expectation value of the Polyakov loop by an 
exponential factor
\beq
<\fhs{-1mm}|\pol|\fhs{-1mm}> \propto \exp \left(-\beta m^{div}   \right) = 
\exp \left(-N_{\tau}am^{div}\right)
\label{eq:eq5Q}
\eeq
depending on the length of the loop $\beta$ times a divergent mass $m^{div}$, 
which is proportional to the ultraviolet cutoff 
\beq
m^{div} \propto \frac{1}{a} \; ,
\label{eq:massdiv}
\eeq
where $a$ is the lattice spacing and ${\beta\fhs{-0.5mm}=\fhs{-0.5mm}N_{\tau} a}$ 
is the inverse temperature. It has been found numerically in~\cite{Dumitru1} 
that the divergent mass term $m^{div}$ is always positive implying that 
$\polAbsAvg$ decreases with finer lattice discretizations 
$N_{\tau}$ along the time direction. 
For this reason the average {\it bare} Polyakov loop actually cannot be 
considered as the caloron holonomy parameter. Since the Polyakov loop is 
renormalizable~\cite{Polyakov1,Gervais2}, it is more appropriate to consider 
a renormalized Polyakov loop independent of $N_{\tau}$.
\begin{figure}[htb]
\centering
\includegraphics[width=0.46\textwidth]{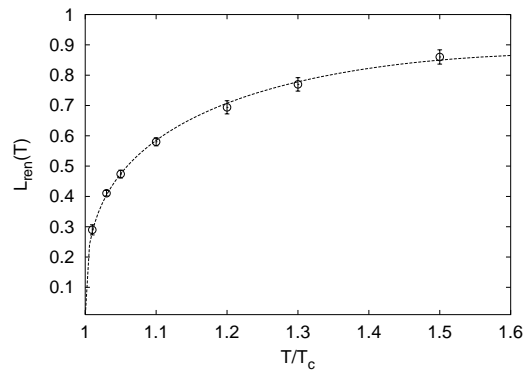}
\caption{The renormalized Polyakov loop in $SU(2)$ LGT versus temperature.
The Figure is taken from Digal \etal~\cite{Digal1}.}
\label{fig:5E}
\end{figure}

A possible renormalization procedure for the Polyakov loop involving its 
spatial correlation function is presented in Refs.~\cite{Digal1,Kaczmarek1}, 
where the renormalized Polyakov loop $L_{\rm ren}(T)$ was defined through the
free energy of a quark-antiquark pair at asymptotic distance. 
The corresponding results are shown in Fig.~\ref{fig:5E}.

The qualitative statement is that the average Polyakov loop is zero in the
confined phase, reflecting the unbroken global $Z(2)$ symmetry, corresponding to a 
caloron ensemble with maximally non-trivial holonomy, while it gradually 
approaches unity above the critical temperature.

\begin{table}[h]
\centering
\begin{tabular}{|c|c|c|}
\hline
T &  $L_{\rm ren}(T)$ & $4\omega(T)$\\ 
\hline
$\le T_c$           &   0.0                               &  1.0         \\ 
$1.10\,T_c$         &   0.58                              &  0.61        \\ 
$1.20\,T_c$         &   0.70                              &  0.51        \\ 
$1.32\,T_c$         &   0.78                              &  0.43        \\ 
$1.54\,T_c$         &   0.85                              &  0.35        \\ 
$\gg T_c$           &   1.0                               &  0.00        \\ 
\hline
\end{tabular}
\caption{Values of the renormalized Polyakov loop as obtained
by Digal \etal~\cite{Digal1}. The holonomy parameter $\omega(T)$ for the 
caloron gas model is fixed (up to the $Z(2)$ symmetry between $\omega$ 
and $\bar\omega$) by the dilute gas relation 
${L_{\rm ren}(T) =  cos(2\pi \omega)}$.}
\label{tab:5A}
\end{table}
Finally, Table \ref{tab:5A} shows for some selected 
temperatures the renormalized Polyakov loop $L_{\rm ren}(T)$ and
the corresponding $\omega$-parameter in the dilute gas approximation.

\subsection{Choice of the caloron density}
\label{subsec:DetOfCalDen}
\vs{1mm}

In lattice QCD Monte Carlo studies directly examining the topological structure
of the gluonic fields one has attempted to get the instanton (or caloron) 
density $n(T)$ by counting the lumps of topological charge. Since such 
investigations in the past required cooling, instantons (calorons) with sizes 
close to the lattice spacing have most likely ``fallen through the grid''.
Furthermore, the sizes of the observed lumps are limited by the lattice 
size. Therefore, the instanton (caloron) densities obtained through this method
are most likely underestimated.

An alternative way to determine the caloron density uses the topological 
susceptibility $\chi$ which can be formally defined in Euclidean space-time by 
\beq
\chi = \int\limits_0^\beta dx_4 \int d^3\vec x <\fhs{-0.5mm}0|
\hat T(q(x)q(0))|0\fhs{-0.5mm}> \;  ,
\label{eq:eq5A}
\eeq
where $q(x)$ is the operator of the topological charge density and 
$\hat T$ the time ordering symbol.

A first estimate of $\chi$ in quenched gauge theory has been obtained from 
$1/N_{\rm color}$ expansion. 
The Witten-Veneziano formula~\cite{Witten1,Veneziano1} relates the topological 
susceptibility of pure gluodynamics to the $\eta'$ mass elevation from the 
masses of the pseudoscalar meson octet. This leads to the prediction 
$\chi\approx (180\, {\rm MeV})^4$. 

The connection to the caloron density $n$ is established through the 
functional integration approach, where (\ref{eq:eq5A}) can be rewritten as
\beq
\chi = \lim\limits_{{V\rightarrow \infty}} \frac{<\fhs{-0.5mm}Q^2\fhs{-0.5mm}>}{V}, 
\quad Q=\int d^4\fhs{-0.5mm}x\, q(x)  \; ,
\label{eq:eq5S}
\eeq
if contact terms are omitted.
The total topological charge can be related to the numbers $N^\pm$ of 
(anti-)calorons in the 4-dimensional volume $V$ by ${Q\fhs{-0.5mm}=
\fhs{-0.5mm}N^+\fhs{-0.5mm}-\fhs{-0.5mm}N^-}$, 
assuming that a collection of these objects exhausts the space-time distribution 
of charge.
For an {\it uncorrelated} caloron gas one would expect a Poisson distribution 
for $N^\pm$ with a variance ${\sigma^2_{N^\pm}\fhs{-1.5mm}=<\fhs{-1.5mm}N^\pm\fhs{-1.5mm}>}$,
leading to the obvious result 
\beq
n = \chi \; ,
\label{eq:eq5D}
\eeq
where ${n\fhs{-0.5mm}=<\fhs{-1.0mm}N^+\fhs{-1.0mm}+
\fhs{-0.5mm}N^-\fhs{-1.0mm}>\fhs{-1.mm}/V}$ 
is the caloron-plus-anticaloron density.

From the renormalization properties of the Yang-Mills theory, however, it 
follows that the caloron numbers are {\it not} Poisson distributed. Instead,
the dispersion of $N^+\fhs{-0.5mm}+\fhs{-0.5mm}N^-$ is significantly smaller, 
${\sigma^2_{N^+\fhs{-0.5mm}+N^-}\fhs{-1.0mm}=
(4/b)\cdot\fhs{-1.0mm}<\fhs{-1.0mm}N^+\fhs{-1.0mm}+\fhs{-0.5mm}N^-\fhs{-1.0mm}>}$, 
where $b=11N_{\rm color}/3$, resulting from some kind of repulsive caloron 
interaction~\cite{ilgenfritz-N,Diakonov3}. 
Assuming that both species, namely calorons and anticalorons, obey this 
dispersion relation independently, being still {\it uncorrelated} among 
each other, one would get the 
modified relation~\cite{ilgenfritz-N}
\beq
n = \frac{b}{4}~\chi
\label{eq:bover4}
\eeq
between density and susceptibility. This would lead to
caloron densities approximately twice as high for 
$SU(2)$ than in the naive approach. However, despite the existence of caloron 
interactions, we will adopt an {\it uncorrelated} sampling of caloron positions 
since the actual correlation between the caloron positions remains undetermined.
Correspondingly, we will adopt relation (\ref{eq:eq5D}) for our {\it uncorrelated}
caloron gas model to fix the density $n(T)$.

Now the temperature dependence of the topological susceptibility shall be discussed. 
Several lattice techniques have been applied to estimate $\chi(T)$. 

The ``cooling method''~\cite{Ilgenfritz1,Teper1} 
evaluates the susceptibility (\ref{eq:eq5S}) by measuring the total charge on 
each independent Monte Carlo configuration by means of an adequate lattice 
operator $q_L(x)$, replacing the continuum operator $q(x)$.  
After some cooling (by local minimization of the action) the (continuum) 
topological charge $Q$ of a configuration can be expressed as 
${Q_L\fhs{-0.5mm}=\fhs{-0.5mm}\sum_x q_L(x)\fhs{-0.5mm}=\fhs{-0.5mm}Z\fhs{-0.5mm}\cdot\fhs{-0.5mm}Q}$ 
with some renormalization constant $Z$. 
In the confined phase the lattice topological charges $Q_L$ converge to 
integer values during the cooling process (${Z\fhs{-0.5mm} \to\fhs{-0.5mm} 1}$), 
but above the critical temperature these plateaux disappear and the method becomes 
ambiguous. Therefore, this method is not adequate to monitor the topological 
susceptibility across the phase transition.

The ``index method''~\cite{Gattringer:2002mr,Teper2,Giusti1} 
determines the topological charge 
${Q\fhs{-0.5mm}=\fhs{-0.5mm}n_L\fhs{-0.5mm}-\fhs{-0.5mm}n_R}$ via the numbers of 
righthanded and lefthanded fermionic zero modes. Only Ginsparg-Wilson 
fermions~\cite{Ginsparg1,GW_rediscovered} are sufficiently 
chiral to allow for an unambiguous 
definition of chiral modes and, consequently, the topological charge.

The ``field theoretical method'' applied in~\cite{Alles1,Alles2} 
evaluates (\ref{eq:eq5A}) by measuring the correlation function directly on the 
lattice, without cooling. In this method, the statistical fluctuations could 
be drastically 
reduced by using improved operators for the topological charge density 
obtained from 1 or 2 smearing steps. The lattice susceptibility $\chi_L$ is 
connected to the continuum susceptibility $\chi$ by
\beq
\chi_L = \langle Q_L^2/ V \rangle = Z^2 a^4 \chi + M 
\label{eq:eq56}
\eeq
with an additional additive renormalization constant $M$ arising from the 
contact terms.

In Ref.~\cite{Alles1,Alles2} the susceptibility $\chi(T)$ for $SU(2)$ has been 
obtained by the field theoretical method. Throughout the confined 
phase, at all ${T\fhs{-0.5mm}<\fhs{-0.5mm}T_c}$, the susceptibility $\chi(T)$ is 
equal to the zero-temperature susceptibility ${\chi(T\fhs{-0.5mm}=\fhs{-0.5mm}0)}$ 
(within error bars). Above the critical temperature the topological susceptibility 
drops to zero. From the 2-smeared results the value 
${\chi(T\fhs{-0.5mm}=\fhs{-0.5mm}0)\fhs{-0.5mm}=\fhs{-0.5mm}(198 \pm 8 {\rm MeV})^4}$ 
is obtained. The caloron densities for ${T\fhs{-0.5mm}>\fhs{-0.5mm}T_c}$, that will be 
adopted for the later calculations, are based on this measurement of $\chi(T)$ and 
are presented in Table \ref{tab:5B}.
\begin{table}[htbp]
\centering
\begin{tabular}{|c|c|}
\hline
T & $n(T)=\chi(T)$ \\ 
\hline
$\le T_c   $     &   $(198\, {\rm MeV})^4$           \\ 
$1.10\, T_c$     &   $(178\, {\rm MeV})^4$           \\ 
$1.20\, T_c$     &   $(174\, {\rm MeV})^4$           \\ 
$1.32\, T_c$     &   $(165\, {\rm MeV})^4$           \\ 
$1.54\, T_c$     &   $(157\, {\rm MeV})^4$           \\ 
$1.79\, T_c$     &   $(136\, {\rm MeV})^4$           \\ 
\hline
\end{tabular}
\caption{Approximate dependence of the caloron density $n(T)$ on the 
temperature $T$ according to ${n(T)\fhs{-0.5mm}=\fhs{-0.5mm}\chi(T)}$ 
from (\ref{eq:eq5D}).
The topological susceptibilities $\chi(T)$ are taken from the $SU(2)$ 
results obtained by Alles et al.~\cite{Alles1} }
\label{tab:5B}
\end{table}

At this point it should be stressed once more, that this estimate of $n(T)$ 
through $\chi(T)$ tacitly assumes that there are topological charge-carrying 
objects (calorons) of charge $\pm 1$, and no correlation exists between the 
placements of single calorons and anticalorons within the volume $V$. If there 
were correlations, the simple estimate of $n(T)$ would no longer hold. 
Assuming that calorons are exhausting the topological structure in the 
deconfined phase, the decline of the topological susceptibility would 
either imply that the caloron density drops to zero or that topologically 
uncharged objects like caloron-anticaloron molecules become dominant in the
deconfined phase, due to an attractive force between oppositely 
charged calorons mediated by the exchange of fermions~\cite{Ilgenfritz2,Shuryak5}. 
 
Since we are dealing with pure Yang-Mills theory, no such correlation will be 
assumed in the following, and the decreasing topological susceptibility 
$\chi(T)$ will be plainly interpreted as a decline of the caloron density $n(T)$. 
However, recent lattice observations~\cite{Ilgenfritz4} have given evidence 
that calorons and their monopole constituents are not the only type of topological 
excitations in the deconfined phase. Instead, non-self-dual magnetic
objects seem to contribute to the vacuum structure at high temperatures, 
above $T_c$. In this respect the model is certainly not complete to describe 
the deconfined phase.

\subsection{Size distributions}
\label{subsec:SizeDis}
\vs{1mm}

At the classical level the action of every single caloron field configuration 
is ${S_{0}\fhs{-0.5mm}=\fhs{-0.5mm}8\pi^2/g^2}$ independent of its collective 
coordinates and therefore all calorons would contribute equally to the functional 
integral. However, the classical scale invariance is broken at the quantum level 
by the quantum weight as introduced in (\ref{eq:oneinstanton}). In the caloron 
model presented here, the quantum weight will be approximately accounted 
for by choosing 
an appropriate size distribution $D(\rho,T)$ according to the single caloron 
quantum weight, which is known in 1-loop order. Since the problem has already 
been encountered in semi-classical simulations based on instantons, let us 
begin with a short discussion of the instanton size distribution. 

\subsubsection{Instanton size distributions}
\label{subsubsec:InstSizeDis}
\vs{1mm}

The calculation of the 1-loop quantum weight in the background of a single 
instanton in $SU(2)$ has been performed by 't Hooft~\cite{tHooft2}
\begin{eqnarray}
\label{eq:eq5U}
Z_1^{\rm{inst}} &=& \int d^4x\, d^3U\, d\rho\, e^{-\frac{8\pi^2}{g^2(\mu)}} f(\rho)\,, \\
f(\rho) &=& \frac{C_0}{4\pi^2}\left(\frac{8\pi^2}{g^2(\mu)}\right)^4 
\frac{1}{\rho^5} (\mu\rho)^b \,, \nonumber \\
b&=&\frac{11}{3}N_{\rm color} - \frac{2}{3}N_f,\quad C_0 \approx 0.64191\,, \nonumber 
\label{eq:tHooft-instdens}
\end{eqnarray}
where the integration $\int d^4x$ is performed over the four-dimensional instanton 
position in space-time and $dU$ denotes the integration over the color group $SU(2)$
according to the Haar measure. Here, $\mu$ is the Pauli-Villars mass, \ie the UV cutoff,
and $g^2(\mu)$ denotes the gauge coupling given at this cutoff. Since the quantum weight is 
proportional to $\rho^{b-5}$, it is divergent in the infrared region for 
${N_{\rm color}\fhs{-0.5mm}\ge\fhs{-0.5mm}2}$. Hence, the integration over the collective 
instanton coordinates diverges unless the large $\rho$ tail is 
suppressed by some other mechanism which would render the model finite.

Several explanations why large instanton sizes could be suppressed in the QCD 
vacuum (confinement effects, higher order effects, repulsive 
instanton interactions) are discussed in~\cite{Shuryak7}. Whereas confinement 
effects were excluded in~\cite{Shuryak7} due to lattice observations, the latter two 
explanations remain viable mechanisms for the suppression of large instantons. 
Assuming the running of the coupling constant $g$ becoming frozen at sufficiently 
large $\rho$ due to higher order effects, one obtains a $1/\rho^5$ 
tail~\cite{Agasian1}. Such a behavior was shown to be not in conflict with 
available lattice data.

A size distribution falling off as ${1/\rho^3}$ would directly yield a linear 
and confining interquark potential with a string tension proportional to the 
coefficient in from of the one-over-cube term~\cite{Diakonov2}, but this could not 
be reconciled with lattice data. 

A third explanation for the large $\rho$ suppression takes the statistical 
mechanics of an interacting instanton gas into account. In~\cite{Ilgenfritz3} 
the large distance interactions between instantons were treated to be of dipole
type as proposed in~\cite{Callan1}. Furthermore, a hard-core type repulsion 
had to be included in order to guarantee the consistency of the calculation.
This finally led to an exponential suppression of large instanton sizes, 
motivating the ansatz for the overall $\rho$-distribution
\beq
D_{inst}(\rho) = a\rho^{b-5} e^{-c\rho^2} \; .
\label{eq:eq5H}
\eeq
The undetermined constants $a$ and $c$ can be fixed by the average instanton 
size and by normalizing the $\rho$ distribution. Exponential suppression 
factors depending on $\rho^2$ have also been obtained in calculations using 
variational techniques~\cite{Diakonov3} and in Monte Carlo 
simulations~\cite{Munster1} taking only hard core type interactions~\cite{Munster2} 
into account. 

\subsubsection{Caloron size distributions}
\label{subsubsec:CalSizeDis}
\vs{1mm}

For KvBLL calorons the calculation of the quantum weight is more difficult
and has been performed only recently. Diakonov et al.~\cite{Diakonov1} found an 
analytical expression for the 1-loop amplitude in the case of an $SU(2)$ caloron.
For the metric in moduli space see also Ref.~\cite{Kraan4}.
In the limit of small $\rho$ again a factor $\rho^{b-5}$ emerges. For 
${\rho\fhs{-0.5mm}\gg\fhs{-0.5mm}\beta\fhs{-0.5mm}=\fhs{-0.5mm}1/T}$ the quantum 
weight becomes 
\begin{eqnarray}
Z_1^{\rm{cal}} &=& \int d^3z_1 \int d^3z_2 T^6 C \left(\frac{8\pi^2}{g^2}  \right)^4
\left(\frac{\Lambda e^{\gamma_E}}{4\pi T}  \right)^\frac{22}{3} \nonumber \\
&\times& \left(\frac{1}{Tr_{12}}  \right)^{\frac{5}{3}} 
\left(vr_{12}+1\right)^{\frac{4v}{3\pi T}-1}  
\left(\bar vr_{12}+1\right)^{\frac{4\bar v}{3\pi T}-1}  \nonumber \\
&\times& \left(2\pi+\frac{v\bar v}{T}r_{12}  
\right) e^{-VP(v) - 2\pi r_{12}P^{''}(v) + ...\;,}  \nonumber
\label{eq:eq5I}
\end{eqnarray}
\beq
P(v) = \frac{1}{12\pi^2 T} v^2\bar{v}^2, \quad P''=\frac{d^2}{dv^2}P(v)  ,
\label{eq:eq5J}
\eeq
where ${C\fhs{-0.5mm}\approx\fhs{-0.5mm} 1.0314}$. Here, the vectors ${\vec z}_1$, ${\vec z}_2$ 
denote the monopole positions, 
${r_{12}\fhs{-0.5mm}=\fhs{-0.5mm}|{\vec z}_1\fhs{-0.5mm}-\fhs{-0.5mm}{\vec z}_2|}$ is the 
monopole separation and ${v\fhs{-0.5mm}=\fhs{-0.5mm}4\pi\omega T}$, 
${\bar v\fhs{-0.5mm}=\fhs{-0.5mm}4\pi\bar\omega T\fhs{-0.5mm}\in\fhs{-0.5mm}[0;2\pi T]}$ 
determine the caloron holonomy. 
$P(v)$ is the 1-loop free energy density for a constant gauge 
field~\cite{Weiss1} corresponding to $v$, $\bar{v}$.
As expected, this formula reduces to the 
quantum weight of the HS caloron~\cite{Gross1} in the case of trivial holonomy. 

For large $\rho$, \ie large $r_{12}$, the quantum weight is dominated by the 
exponential expression. The polynomial $P''(v)$, appearing in the exponent,
is of second order and its zeroes are 
${v_{\pm} =\pi T(1 \pm 1 / \sqrt{3})}$. 
Therefore, large $\rho$ parameters are suppressed by the quantum weight as long as 
${v\fhs{-0.5mm}<\fhs{-0.5mm}v_-}$ or ${v\fhs{-0.5mm}>\fhs{-0.5mm}v_+}$ corresponding 
to nearly trivial holonomy, whereas in the opposite case the quantum weight diverges 
exponentially for large $\rho$. In the latter case a mechanism to cut off the 
$\rho$-distribution is needed to guarantee that the single caloron remains
an important part of the vacuum structure.

For small $\rho$, the exponential suppression ansatz 
(\ref{eq:eq5H}) reflecting instanton interactions could also be applied to calorons. 
Since this ansatz has been deduced for instantons interacting like pointlike 
dipoles, this is a rather crude parametrization for the caloron $\rho$-distribution 
because of its more complex structure with two emerging monopoles.
Additionally, the $\rho$ parameter, which serves as a good approximation for 
the four dimensional radius of the instanton action lump, does not describe 
the action lump size in the case of a dissociated caloron. However, for 
${\rho\fhs{-0.5mm}\ll\fhs{-0.5mm}\beta}$ the caloron is not dissociated and the 
two monopoles are merged together in one action lump of nearly spherical shape. 
In this limit $\rho$ can justifiably be interpreted as caloron size. Then it is 
reasonable to take over the distribution (\ref{eq:eq5H}) for sufficiently small 
calorons, provided that the majority of the calorons is not dissociated. 

In a lattice study~\cite{Chu2} searching for HS calorons at finite 
temperatures, the temperature dependent average caloron size $\bar\rho(T)$ 
has been measured by comparing the topological charge density two-point 
correlator with that of a superposition of single HS caloron 
profiles. The conclusion was that in the confined phase the average caloron 
size as a function of temperature is approximately constant, 
${\bar\rho(T\fhs{-0.5mm}<\fhs{-0.5mm}T_c)\fhs{-0.5mm}=\fhs{-0.5mm}0.33}$~fm, 
whereas it falls off in the deconfined phase. Therefore, the same exponential 
suppression factor $\exp(-c\rho^2)$ can be applied to calorons at sufficiently 
small temperatures $T$ due to ${\bar\rho(T)/\beta \ll 1}$.

\begin{table*}[htp]
\centering
\begin{tabular}{|l|l|l|}
\hline
Temperature & \hs{17mm}$\rho$-distribution & \hs{6mm}fixing of coefficients by \\ 
\hline
$T<T_c$ & $D(\rho,T) = A \cdot \rho^{b-5}\cdot \exp(-c\rho^2)$ & $\int D(\rho,T) 
d\rho=1,\quad \bar\rho$ fixed \\
$T>T_c$ & $D(\rho,T) = A \cdot \rho^{b-5}\cdot \exp(-\frac{4}{3}(\pi\rho T)^2)$ & 
$\int D(\rho,T) d\rho=1,\quad \bar{\rho}$ running\\
\hline
\end{tabular}
\caption{The $\rho$-distributions used in the confined and deconfined phase. 
In the confined phase the coefficients $A$ and $c$ are fixed by reproducing
a prescribed average caloron size and the normalization of 
$D(\rho,T)$. In the deconfined phase there is only one coefficient, $A$, 
fixed by normalization.}
\label{tab:5C}
\end{table*}

However, this approximation is only valid up to a certain temperature 
${T^{*}\fhs{-0.5mm}<\fhs{-0.5mm}T_c}$ where $\beta$ becomes of the same 
order as $\bar\rho$. 
Above this temperature $T^{*}$ the application of the 
simple instanton-radius 
suppression formula to the (already partly dissociated) calorons is 
no longer justified. Instead, one can consider the statistical mechanics of 
an interacting monopole gas in a similar manner. 
Effectively, this can be parametrized by the quantity $\rho_{3D}$, 
which is meant to coincide with $\rho$ at ${\rho\x\ll\x\beta}$ (then describing
the radius of the $O(4)$-symmetric, non-dissociated caloron), and to provide
the 3-dimensional radius of the caloron constituent monopoles (which have identical 
shape for ${\omega\x=\x 0.25}$) at ${\rho\x\gg\x\beta}$. Connecting these two limits continuously,
${\rho_{3D}(\rho)}$ is a function of the caloron parameter $\rho$. 
In terms of this quantity the statistical mechanics of an interacting monopole gas 
can effectively be described by the ansatz
\beq
D_1(\rho,T) d\rho = D_{inst}(\rho_{3D}) d\rho_{3D}
\label{eq:rhoDisCal1}
\eeq
which should lead to a distribution of the caloron ``size'' parameter $\rho$.

\begin{figure}[h]
\centering
\includegraphics[angle=0,width=0.46\textwidth]{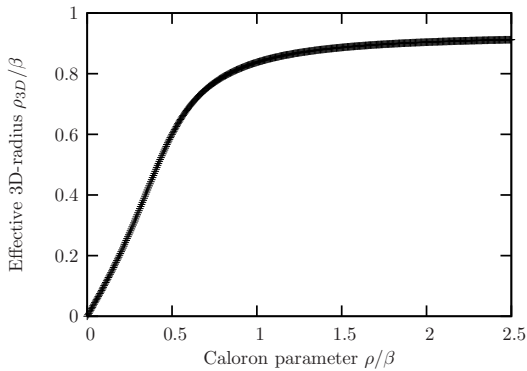}
\caption{Dependence of the effective 3-dimensional caloron/monopole radius 
$\rho_{3D}$ on the caloron parameter $\rho$. 
Here, $\rho_{3D}$ is quantitatively defined in the text as the effective,
3-dimensional radius of the region where the action density exceeds a certain 
fraction of the maximal density in the caloron/monopole center. This fraction
is chosen such, that ${\rho_{3D}\x=\x \rho}$ at ${\rho\x\ll\x\beta}$.}
\label{fig:rho3D}
\end{figure}

Here, we are using a numerically accessible definition meeting the above 
requirements. We define $\rho_{3D}$ as the effective 3-dimensional radius 
of a region where the action density of a caloron (or of its monopole
constituents, respectively) exceeds a certain fraction of the maximal density 
in the caloron (or monopole) center. Its functional dependence on $\rho$ 
is shown in Fig.~\ref{fig:rho3D}.
The precise cutoff fraction is chosen such that 
${\rho_{3D}\fhs{-0.5mm}\approx\fhs{-0.5mm}\rho}$ 
for ${\rho\fhs{-0.5mm}\ll\fhs{-0.5mm}\beta}$. 
Hence, the ansatz (\ref{eq:rhoDisCal1}) becomes 
equivalent to the instanton distribution $D_{inst}(\rho)$ 
as long as ${\bar \rho\fhs{-0.5mm}\ll\fhs{-0.5mm}\beta}$. 
It also remains applicable at high temperatures. However, with increasing 
temperature the extent of the monopoles $\rho_{3D}$ is kinematically bounded 
by $\beta/\omega$. 
Therefore, 
this application of (\ref{eq:eq5H}) to the interacting monopole gas alone 
cannot yield a cut off for the $\rho$-distribution. 

What is needed in order to finally cut-off the $\rho$-distribution
is a suppression of large distances $d$ between the constituents.
Such an additional suppression factor for large ``size'' parameters 
${\rho\fhs{-0.5mm}\gg\fhs{-0.5mm}\beta}$ can be obtained from
the quantum weight (\ref{eq:eq5I}). In~\cite{Hofmann1} the 3-dimensional 
volume $V$ in the quantum weight has been interpreted as a specific caloron volume
(elementary cell per caloron in a dense packing) $V_{Cal}(\rho)$ 
depending on the size parameter $\rho$. 
We introduce a single caloron volume, considered 
in the limiting case of a well dissociated caloron and dense packing as the 
volume of a cylinder, circumscribing the caloron
\beq
V_{Cal}(\rho)=C_0(\omega,T)\pi\beta^2d, \quad d = \frac{\pi\rho^2}{\beta} \; .
\label{eq:VolCal}
\eeq
Here ${\beta\fhs{-0.5mm}=\fhs{-0.5mm}1/T}$ sets the scale of the monopole size, 
and $d$ is the monopole separation. $C_0(\omega,T)$ is a holonomy dependent, 
undetermined factor of order unity. 
Substituting the volume $V$ by this ansatz for the specific 
caloron volume $V_{Cal}(\rho)$ in the quantum weight (\ref{eq:eq5I}) and 
neglecting all non-exponential terms one arrives at the form
\beq
D_2(\rho,T) \stackrel{\rho\rightarrow\infty}{\sim} e^{-\frac{4}{3}
(\pi\rho T)^2G(\omega)}
\label{eq:second}
\eeq
with
\beq
G(\omega)= 16C_0(\omega)\pi^2\omega^2\bar\omega^2  
         + 4 (\omega^2 + \bar\omega^2 - 4\omega\bar\omega)  
\label{eq:eq5M}
\eeq
for the large $\rho$ suppression. For trivial holonomy the polynomial 
$G(\omega)$ becomes unity, and the well known temperature suppression for the 
HS calorons~\cite{Gross1} is recovered, but for non-trivial holonomy the 
disastrous repulsive (positive) coefficient of $d$ 
appearing in the exponential is now overruled by the specific caloron 
volume. Choosing ${C_0(\omega,T)\fhs{-0.5mm}>\fhs{-0.5mm}0.82}$ is 
plausible since $C_0(\omega,T)$ was assumed to be of order unity. This is sufficient 
to yield an exponentially suppressed $\rho$-distribution even though in this case 
no caloron interaction (other than the excluded volume interaction) was taken 
into account. In principle, the coefficient $C_0$ has to be determined in 
a self-consistent way, such that the sum of all specific caloron volumes 
$V_{Cal}$ equals the total volume $V$ up to a factor close to unity.

The strength of this suppression rises with the temperature and depends 
on the holonomy. Eq. (\ref{eq:VolCal}), approximating the specific 
caloron volume as a cylinder, certainly does not hold for the spherical
HS calorons well above $T_c$.
Therefore, this ansatz is assumed to describe the $\rho$-distribution at 
high temperatures but still in the confined phase. 

\begin{figure}[h]
\centering
\includegraphics[angle=0,width=0.46\textwidth]{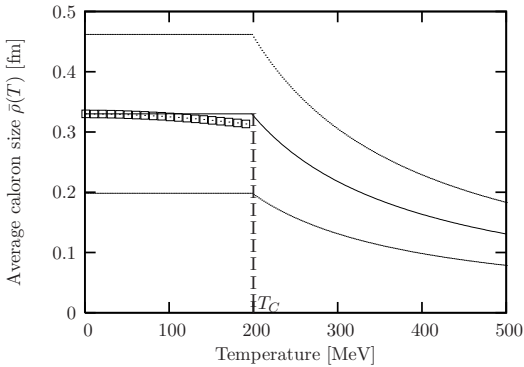}
\caption{Illustration of the dependence of the average caloron size $\bar\rho$ 
(solid) and its interval $[\bar\rho-\Delta_{\bar\rho},\bar\rho-\Delta_{\bar\rho}]$ 
of (standard) deviation (dotted) on temperature $T$. The two different ansatzes 
for the $\rho$-distributions for the confined/deconfined phase are continuously 
connected at the critical temperature $T_c$.
For the curve indicated by open squares see the text.}
\label{fig:5I}
\end{figure}

Sewing together both ansatzes corresponding to the opposite temperature 
regimes ${\bar\rho\fhs{-0.5mm}\ll\fhs{-0.5mm}\beta}$ and 
${\bar\rho\fhs{-0.5mm}\gg\fhs{-0.5mm}\beta}$ in the simplest form, 
leads to the ansatz
\beq
D(\rho,T) =  a\rho_{3D}^{b-5} e^{-c\rho_{3D}^2} \cdot \frac{d\rho_{3D}}{d\rho} 
\cdot e^{-\frac{4}{3}(\pi\rho T)^2G(\omega)}
\label{eq:RhoDisCalFinal}
\eeq
for the $\rho$-distribution in the confined phase. This combination of both 
approaches guarantees that the average caloron size is bounded from above 
for all temperatures. To show this, we plot the average caloron size 
${\bar\rho(T\fhs{-0.5mm}<\fhs{-0.5mm}T_C)}$ obtained by this ansatz (\ref{eq:RhoDisCalFinal})
for the plausible scenario ${\omega\fhs{-0.5mm}=\fhs{-0.5mm}0.25}$, 
${C_0\fhs{-0.5mm}=\fhs{-0.5mm}1}$ in Fig.~\ref{fig:5I} (square symbols). 
Here $c$ was determined such that at zero temperature 
${\bar\rho(T=0) = 0.33}$~fm.

We see, that through this rough ansatz (\ref{eq:RhoDisCalFinal}) the average
of $\rho$ at any temperature throughout the confined phase,
${\bar\rho(T\fhs{-0.5mm}<\fhs{-0.5mm}T_C)}$, 
is not only bounded but also approximately constant 
as has been observed in lattice studies~\cite{Chu2}. 
Here it is due to the weakened explicit temperature dependence for $\omega=0.25$.

Since the arguments were rather crude and the coefficient $C_0$ remains
undetermined, only the ubiquitous form of an exponential suppression factor
$\exp(-c\rho^2)$ will be subscribed for our model, while its actual coefficient
$c$ should be fixed according to lattice observations. In principle, these
can be obtained by comparing lattice measurements of the topological
density correlator with model calculations using the KvBLL caloron profile,
analogous to the determination of
${\bar\rho(T\fhs{-0.5mm}<\fhs{-0.5mm}T_C)}$ performed in~\cite{Chu2}
which were based on the HS caloron profile.
In the following subsection we will find an alternative prescription
to fix ${\bar\rho(T\fhs{-0.5mm}<\fhs{-0.5mm}T_C)}$ 
based on the space-like string tension leading to a 
self-consistent tuning for our model.

In the deconfined phase the caloron holonomy ceases to be maximally non-trivial
and approaches its trivial value as discussed in Section~\ref{subsec:DetOfHol}. 
Therefore, 
the HS caloron quantum weight, which converges to zero for 
${\rho\fhs{-0.5mm}\rightarrow\fhs{-0.5mm}\infty}$, 
explicitly determines $D(\rho,T)$ for ${T\fhs{-0.5mm}\gg\fhs{-0.5mm}T_c}$. Since 
the caloron density $n(T)$ is assumed to fall off fast beyond the phase transition 
as argued in Section~\ref{subsec:DetOfCalDen}, caloron interactions do not need 
to be considered. Therefore, the coefficient of $\rho^2$ in the exponential is 
exactly known (set by the temperature) and the average caloron size arises directly 
from (\ref{eq:eq5I}). 

Hence, we are left with two different size distributions for the confined and the
deconfined phase, as summarized in Table~\ref{tab:5C}. For our model, we sew together
these two ansatzes for the $\rho$-distribution in a continuous manner, \ie
\beq
\bar\rho(T_C)_{conf.} = \bar\rho(T_C)_{deconf.}
\label{eq:FixOfTC}
\eeq
which is shown in Fig.~\ref{fig:5I} for the case 
${\bar\rho(T<T_C)=0.33}$~fm. 
We will identify the temperature, at which both ansatzes are continuously 
connected, as the critical temperature $T_C$. This establishes a relation between 
${\bar\rho(T\fhs{-0.5mm}<\fhs{-0.5mm}T_C)}$ and $T_C$ that can be used to fix 
${\bar\rho(T\fhs{-0.5mm}<\fhs{-0.5mm}T_C)}$ in a consistent and unambiguous manner 
{\it within} our model.

\subsubsection{Final determination of ${\bar\rho(T\fhs{-0.5mm}<\fhs{-0.5mm}T_C)}$}
\label{subsubsec:FinalFixRho}
\vs{1mm}

From lattice calculations~\cite{Lucini1}, the zero-temperature string tension 
${\sigma(T\fhs{-0.5mm}=\fhs{-0.5mm}0)}$ in terms of the critical temperature is 
very accurately known for $SU(2)$ to be 
${T_C/\sqrt{\sigma(T\fhs{-0.5mm}=\fhs{-0.5mm}0)}\fhs{-0.5mm}=\fhs{-0.5mm}0.709\,(4).}$ 
Furthermore, it is well known that the space-like string tension $\sigma_S(T)$ 
coincides with the real (time-like) string tension at 
${T\fhs{-0.5mm}=\fhs{-0.5mm}0}$, but stays approximately constant for all 
temperatures in the confined phase before it even 
rises quadratically with $T$ above the critical temperature~\cite{Lucini1}. Hence, 
we obtain the relation for the space-like string tension
\vs{-1.5mm}
\beq
\label{eq:StringTensionTC}
\sigma_S(T<T_C) \approx \left( \frac{T_C}{0.71} \right)^2 \; .
\eeq
\vs{-0.5mm}
Together with the relation between $T_C$ and 
${\bar\rho(T\fhs{-0.5mm}<\fhs{-0.5mm}T_C)}$ 
imposed by (\ref{eq:FixOfTC}), one obtains a very precise determination of 
${\bar\rho(T\fhs{-0.5mm}<\fhs{-0.5mm}T_C)}$, since the space-like string tension 
$\sigma_S(T)$ is very sensitive to the average caloron size parameter $\bar\rho(T)$. 

\begin{figure}[h]
\centering
\includegraphics[angle=0,width=0.46\textwidth]{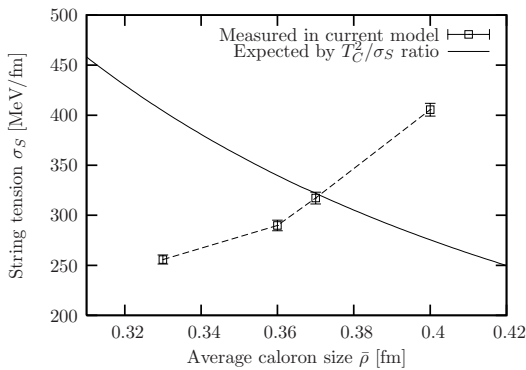}
\caption{Space-like string tensions obtained from our model for some selected 
values of ${\bar\rho(T\fhs{-0.5mm}<\fhs{-0.5mm}T_C)}$ together with the results 
for $\sigma_S(T_C)$ derived from (\ref{eq:FixOfTC}) and (\ref{eq:StringTensionTC}).}
\label{fig:5K}
\end{figure}

Fig.~\ref{fig:5K} shows the space-like string tension obtained from our model, 
as will be described in detail in the next Section, for some selected values of 
${\bar\rho(T\fhs{-0.5mm}<\fhs{-0.5mm}T_C)}$
together with those results for $\sigma_S(T_C)$, that are derived from 
(\ref{eq:FixOfTC}) and (\ref{eq:StringTensionTC}). 
This approach unambiguously fixes the average caloron size to 
${\bar\rho(T\fhs{-0.5mm}<\fhs{-0.5mm}T_C)\fhs{-0.5mm}=\fhs{-0.5mm}0.37}$~fm. 
It should be noticed that this result is not only consistent
with the lattice observations for the space-like string tension~\cite{Lucini1}, 
but also with the direct lattice studies for the caloron size~\cite{Chu2}, 
which gave 
${\bar\rho(T\fhs{-0.5mm}<\fhs{-0.5mm}T_C)\fhs{-0.5mm}=\fhs{-0.5mm}0.33}$~fm. 
The small deviation of approximately $10\%$ can easily be explained by the fact 
that the average caloron size in the lattice studies was extracted by assuming 
the underlying topological objects to be HS-calorons. It would be worthwhile
to repeat the calculation of the shape of the topological density correlator
for uncorrelated KvBLL calorons and to compare with the data.

As a byproduct of the determination of ${\bar\rho(T\fhs{-0.5mm}<\fhs{-0.5mm}T_C)}$
we have also fixed the critical temperature 
${T_C\fhs{-0.5mm}\approx\fhs{-0.5mm} 178}$~MeV
as well as the zero-temperature string tension 
${\sigma(0)\fhs{-0.5mm}\approx\fhs{-0.5mm}318}$~MeV/fm {\it within} our model.

\section{Results for interquark potentials}
\label{sec:Confinementresults}

\begin{table*}[htb]
\centering
\begin{tabular}{|c|c|c|c|c|c|c|c|c|c|}
\hline
$T$ & $N_s^3\times N_{\tau}$ & $n^{\frac{1}{4}}\,[{\rm MeV}]$ & $4\omega$ & 
$\bar\rho\,[{\rm fm}]$& $\bar\rho/a$ & \# & $cos(2\pi\omega)$& $<\fhs{-1mm}|\pol|\fhs{-1mm}>$ 
&  $\gamma$  \\ 
\hline
$0.80\, T_C$  & $32^3\times 10$& 198 & 1.00 & $ 0.37 $   
		& 2.66 & 777 & 0.00  & $0.13 \pm 0.01$ & $1.61 \pm 0.01$ \\
$0.90\, T_C$  & $32^3\times 9$ & 198 & 1.00 & $ 0.37 $   
		& 2.69 & 591 & 0.00  & $0.14 \pm 0.01$ & $1.65 \pm 0.01$ \\
$1.00\, T_C$  & $32^3\times 8$ & 198 & 1.00 & $ 0.37 $   
		& 2.67 & 526 & 0.00  & $0.14 \pm 0.01$ & $1.69 \pm 0.01$ \\ 
\hline
$1.10\, T_C$  & $32^3\times 8$ & 178 & 0.61 & $ 0.33 $   
		& 2.62 & 160 & 0.58  & $0.43 \pm 0.01$ & $1.29 \pm 0.01$ \\
$1.20\, T_C$  & $32^3\times 8$ & 174 & 0.51 & $ 0.31 $   
		& 2.69 & 160 & 0.70  & $0.59 \pm 0.01$ & $1.18 \pm 0.01$ \\
$1.32\, T_C$  & $32^3\times 8$ & 165 & 0.43 & $ 0.28 $   
		& 2.66 & 160 & 0.78  & $0.72 \pm 0.01$ & $1.10 \pm 0.01$ \\
\hline
\end{tabular}
\caption{For the selected temperatures 
${T/T_C\fhs{-0.5mm}=\fhs{-0.5mm}0.8,0.9,1.0,1.10,1.20,1.32}$ 
the corresponding model parameters 
$n(T),\,\omega(T),\,\bar\rho(T)$, chosen according to 
Section~\ref{sec:Inputparameters}, 
as well as the number of configurations \#, the lattice sizes 
${N_s\fhs{-0.5mm}\times\fhs{-0.5mm} N_{\tau}}$, 
and the average caloron size in lattice units $\bar\rho/a$ are listed. 
Furthermore, the measured Polyakov loop 
$<\fhs{-1mm}|\pol|\fhs{-1mm}>$ together with the input value value 
${cos(2\pi\omega)}$ and the action surplus factor $\gamma$ as defined 
in (\ref{eq:DefActionSurplus}) are shown.}
\label{tab:6E}
\end{table*}

Let us now examine the confinement properties of our model at several temperatures. 
For this purpose, caloron ensembles corresponding to temperatures 
${T/T_C\fhs{-0.5mm}=\fhs{-0.5mm}0.8,\, 0.9,\, 1.0}$ 
for the confined and ${T/T_C\fhs{-0.5mm}=\fhs{-0.5mm}1.10,\, 1.20,\, 1.32}$ for the deconfined 
phase have been generated and discretized on suitable lattices. 
As in Section~\ref{sec:holonomy_decides}, the lattices do not obey 
any boundary condition. The performed calculations 
are {\it open volume} simulations, since this is the easiest way to guarantee that 
even strongly dissociated calorons would fit on the lattice. The lattice spacing 
$a$ was chosen such, that the ratio $\bar\rho(T)/a$ stays approximately constant 
for the selected temperatures. The lattice sizes are listed in Table \ref{tab:6E} 
together with all other relevant model parameters, which are chosen 
according to Section~\ref{sec:Inputparameters}. Table \ref{tab:6E} also shows the 
measured {\it modulus} of the average Polyakov loop $<\fhs{-1mm}|\pol|\fhs{-1mm}>$ 
together with the input value of the average Polyakov loop ${cos(2\pi\omega)}$,
as well as the action surplus factor $\gamma$ as defined in 
(\ref{eq:DefActionSurplus}). The amount of $60 \ldots 70 \%$ additional action
in the confinement phase seems to be acceptable.

We begin with the discussion of the space-like string tension before 
turning to the more interesting time-like potentials that will be derived 
from the Polyakov loop correlator. As seen in Section~\ref{subsubsec:FinalFixRho}
the space-like string tension is a sensible tool to determine the caloron 
size parameter.

\subsection{Spatial Wilson loops}
\label{subsec:wilson}
\vs{1mm}

The color singlet ground state energy of a heavy quark-antiquark 
pair can be extracted from the Wilson loop
\beq
W(C_{R,R_T}) = \frac{1}{N_C}\; Tr \prod\limits_{l\in C_{R,R_T}} U_l
\label{eq:wilsonloop}
\eeq
in the limit $R_T\rightarrow\infty$ according to
\beq
E(R) = -\fhs{-2mm}\lim\limits_{R_T\rightarrow\infty} \frac{1}{R_T}\, \ln <W(C_{R,R_T})> + C_0\,,
\label{eq:eqPub1}
\eeq
where $C_{R,R_T}$ denotes a rectangular path with spatial and temporal 
extent $R$ and $R_T$, respectively.
However, at finite temperature this limit can not be realized due to 
the compactification of time. Instead, {\it spatial} Wilson 
loops $W(C_{R,R_2})$ and their associated potential will be considered here. 
It is supposed to coincide with the physical quark 
antiquark potential at sufficiently low temperatures due to the effective 
isotropy of space-time at 
${T\fhs{-0.5mm}=\fhs{-0.5mm}0}$. For higher temperatures, spatial Wilson 
loops do not provide a 
physical potential. This is obvious, since the spatial string tension stays 
approximately 
constant in the confined phase and even rises beyond the critical temperature.

\begin{figure}[htb]
\centering
\includegraphics[angle=0,width=0.48\textwidth]{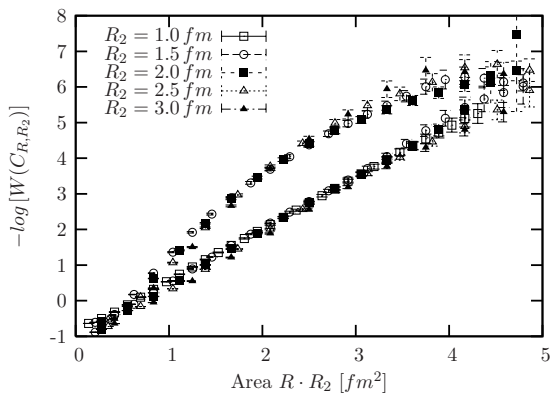}
\caption{Negative logarithm of rectangular Wilson loops, $-log(W(C_{R,R_2}))$, 
with side lengths $R,\,R_2$ in fundamental (lower curve) and adjoint representation versus
the area ${R\fhs{-0.5mm}\cdot\fhs{-0.5mm}R_2}$. The different symbols correspond to different 
side lengths $R_2$.}
\label{fig:60A}
\end{figure}

First of all, we check whether the spatial Wilson loops fulfill the area law
\beq
\label{eq:AreaLaw}
<W(C_{R,R_2})> \propto e^{-\sigma\cdot R\cdot R_2} \; .
\eeq
We show in Fig.~\ref{fig:60A} the negative logarithm of the Wilson loops 
$<W(C_{R,R_2})>$ as a function of the enclosed area 
${R\fhs{-0.5mm}\cdot\fhs{-0.5mm} R_2}$ of the loops. 
For the fundamental representation (lower curve) the dependence 
is almost linear as expected, 
except for some loops with strongly deformed geometry, 
\ie ${~R_2\fhs{-0.5mm}\gg\fhs{-0.5mm} R}$. 
This deviation is due to non-area terms additionally which have been omitted in the 
exponent of (\ref{eq:AreaLaw}). 
For the adjoint representation, however, the curve also starts rising 
linearly at small 
distances but flattens off at larger $R$. This can be seen as a 
first evidence that screening effects for the adjoint charges can be 
reproduced by our model. 
\begin{figure*}[htb]
\centering
\begin{tabular}{cc}
\hs{6mm}(a) &\hs{6mm}(b) \\
\includegraphics[angle=0,width=0.48\textwidth]{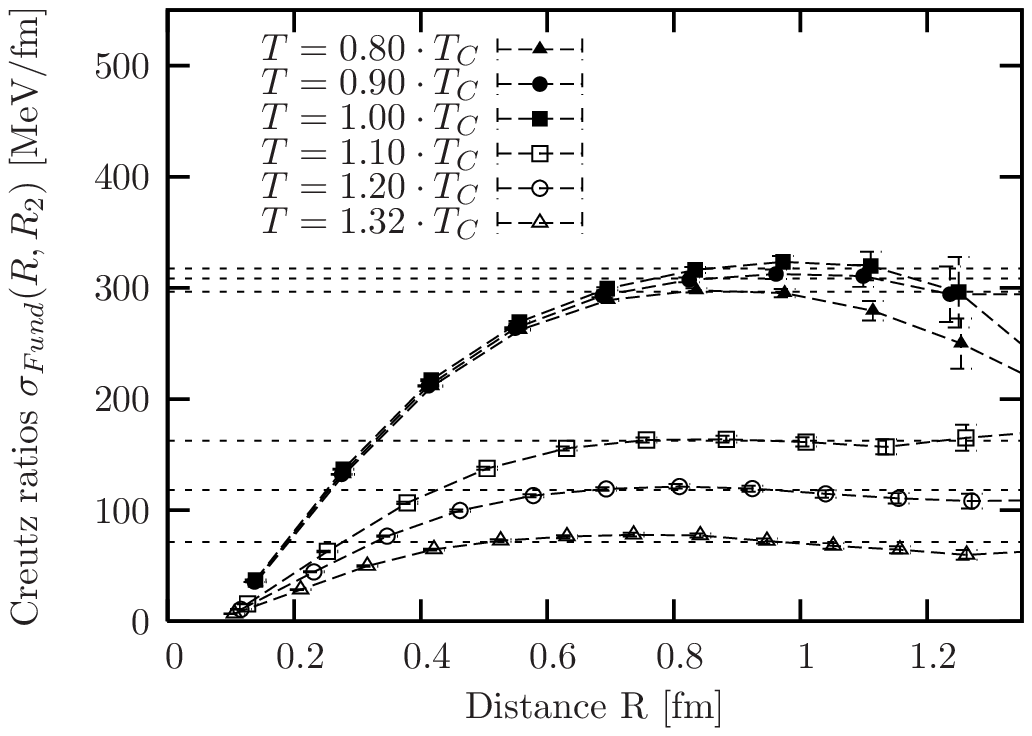}   &   
\includegraphics[angle=0,width=0.48\textwidth]{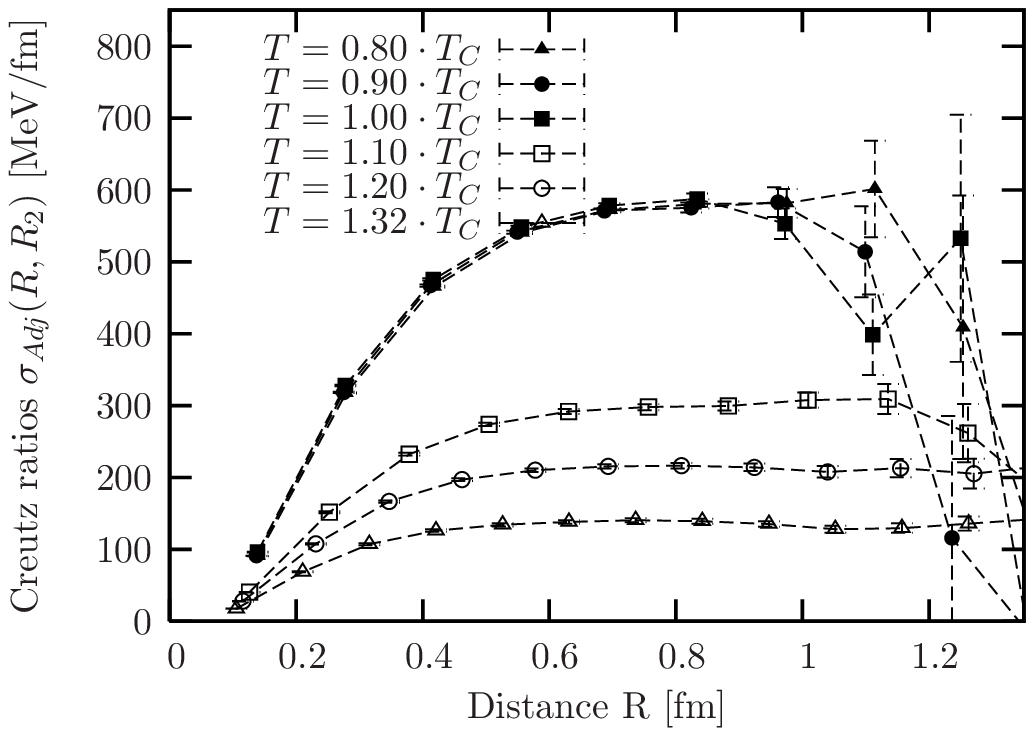}\\  
\end{tabular}
\caption {Effective string tension $\sigma(R,R_2)$ calculated by Creutz 
ratios (with ${R_2\fhs{-0.5mm}=\fhs{-0.5mm}2\fhs{-0.5mm}\cdot\fhs{-0.5mm} R}$) 
from spatial Wilson loops versus the distance $R$ at different temperatures 
${T/T_C\fhs{-0.5mm}=\fhs{-0.5mm}0.8,\, 0.9,\, 1.0}$ for the confined and at 
${T/T_C\fhs{-0.5mm}=\fhs{-0.5mm}1.10,\, 1.20,\, 1.32}$ for the deconfined phase
in the fundamental~(a) and in the adjoint~(b) representation.}
\label{fig:60G}
\end{figure*}

There are various sources for non-area terms to appear in the exponent 
of Eq. (\ref{eq:AreaLaw}). In general, the effect of various obscuring non-area 
terms can be compensated by extracting the
string tension from certain products of loops of similar shape. 
A perimeter term, for example, is caused by 
the self-energy contribution, which is proportional to the perimeter 
${R\fhs{-0.5mm}+\fhs{-0.5mm}R_2}$, 
leading to 
${<W(C_{R,R_2})> \propto\fhs{-0.5mm}
 \exp({-\sigma RR_2 - \alpha(R\fhs{-0.5mm}+\fhs{-0.5mm}R_2) - \gamma})}$. 
The constant and perimeter terms can be eliminated by calculating 
the effective string tension from the Creutz ratios~\cite{Creutz1}
\begin{eqnarray}
\sigma(R,R_2)  = &  \\
\frac{1}{2a^2} \times \ln & 
\left[ \frac{<W(C_{R+1,R_2-1})><W(C_{R-1,R_2+1})>}{<W(C_{R,R_2})><W(C_{R,R_2})>}\right] \; .
\nonumber 
\label{eq:eq2ZZI}
\end{eqnarray}
However, while the extent $R_2$ of the contour $C_{R,R_2}$ should be 
sufficiently large 
to suppress contributions of excited states, the expectation value of 
the Wilson loop increasingly
suffers from statistical fluctuations with growing $R_2$. Hence, a 
reasonable value for the side 
length $R_2$ has to be chosen to evaluate the effective string tension. 
As a compromise between 
systematical and statistical errors we choose the side length $R_2$ 
twice as long as $R$, \ie 
${R_2\fhs{-0.5mm}=\fhs{-0.5mm}2\fhs{-0.5mm}\cdot\fhs{-0.5mm}R}$, 
which is widely seen as a valid, albeit minimal, condition.

The resulting effective string tensions obtained from Creutz 
ratios with ${R_2\fhs{-0.5mm}=\fhs{-0.5mm}2\fhs{-0.5mm}\cdot\fhs{-0.5mm} R}$ 
are presented in Fig.~\ref{fig:60G}a for the fundamental 
representation. For all  
temperatures shown below $T_C$ they run into a plateau of 
approximately ${296-318}$~MeV/fm within 
the error bars at moderate distances of about 
${R\fhs{-0.5mm}\approx\fhs{-0.5mm} 0.8}$~fm.
For fixed $\bar\rho(T)$ the observed space-like string 
tension does only depend weakly on temperature.
An exact temperature independence could easily be obtained by a 
minimal change of the
average caloron size due to its strong effect on the space-like 
string tension.
For the presented temperatures corresponding to the deconfined 
phase the spatial string
tension runs into plateaux at somewhat smaller distances. 
The corresponding values of the string tension are
strongly reduced compared to the confined phase. 
This is due to the decreased caloron density, which was 
chosen according to the drop of the topological susceptibility 
$\chi(T)$ in the deconfined phase. This is a 
limitation of the current model focussing on
(anti-)self-dual monopoles as the exclusive origin of the spatial string tension.

All curves, however, tend to fall off as $R$ increases. 
If one takes this noisy observation serious it would be
interesting to study the contribution of caloron systems 
to very large Wilson loops analytically. Since the
caloron gauge fields become Abelian far outside their cores, 
it is even worthwhile and instructive to consider only the 
contributions of the (Abelian) caloron far fields of a caloron 
gas to very large Wilson loops. 
Such analytical calculations are currently under consideration~\cite{Bruckmann3}.

Fig.~\ref{fig:60G}b shows the effective adjoint string tensions, which are 
expected to run into plateaux before falling off 
due to adjoint charges being screened by gluons. 
In our model, in the confined phase the adjoint string tensions reach plateaux 
of $580$~MeV/fm at a distance of $R\approx 0.8$~fm. At larger 
distances the color screening sets in, 
forcing the adjoint force to fall off. We could clearly observe the takeover
by screening effects in the confined phase at distances ${R\x\approx\x 1}$~fm when 
considering ``square'' Creutz ratios with ${R_2\x=\x R}$. However, 
for the extended Creutz ratios with ${R_2\x=\x 2R}$ the signal becomes very noisy at 
${R\x>\x 1}$~fm, and hence, we can only see slight indications for 
screening effects here. For the deconfined phase the adjoint 
forces are much smaller than below the critical temperature,  
due to the reduced caloron density.

\begin{figure}[htb]
\centering
\includegraphics[angle=0,width=0.48\textwidth]{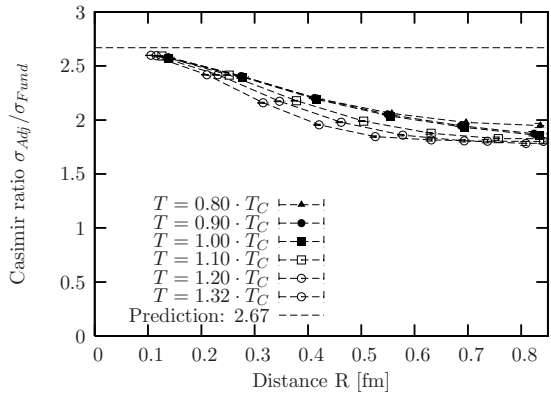}
\caption{Casimir ratio $\sigma_{Adj}/\sigma_{Fund}$ of adjoint and fundamental 
forces from space-like Wilson loops versus the distance $R$
at the different temperatures ${T/T_C\fhs{-0.5mm}=\fhs{-0.5mm}0.8,\, 0.9,\, 1.0}$ 
for the confined phase and at 
${T/T_C\fhs{-0.5mm}=\fhs{-0.5mm}1.10,\, 1.20,\, 1.32}$ for the deconfined phase.
The theoretical prediction of Casimir scaling is marked by the horizontal 
dashed line.}
\label{fig:60H}
\end{figure}

Finally, the Casimir scaling hypothesis can be checked. 
While the scaling of the forces (according to the quadratic Casimir operators)
between sources in varios representations of the color group is primarily a
property of the one-gluon exchange, it holds surprisingly well also at intermediate
distances, thus constraining the confining mechanisms. It was an argument against 
the instanton forces that the Casimir scaling would be strongly 
violated~\cite{Shevchenko}.
The ratio ${\sigma^{adj}(R)/\sigma^{fund}(R)}$ obtained in our model
is plotted versus the quark separation $R$ in Fig.~\ref{fig:60H}.
It can be seen that the ratio of adjoint and fundamental 
effective string tensions begins very close to the predicted value at small 
distances. 

\subsection{Polyakov loops}
\label{subsec:polyakov}
\vs{1mm}

\begin{figure*}[htb]
\centering
\begin{tabular}{cc}
\hs{6mm}(a) & \hs{6mm}(b) \\
\includegraphics[angle=0,width=0.48\textwidth]{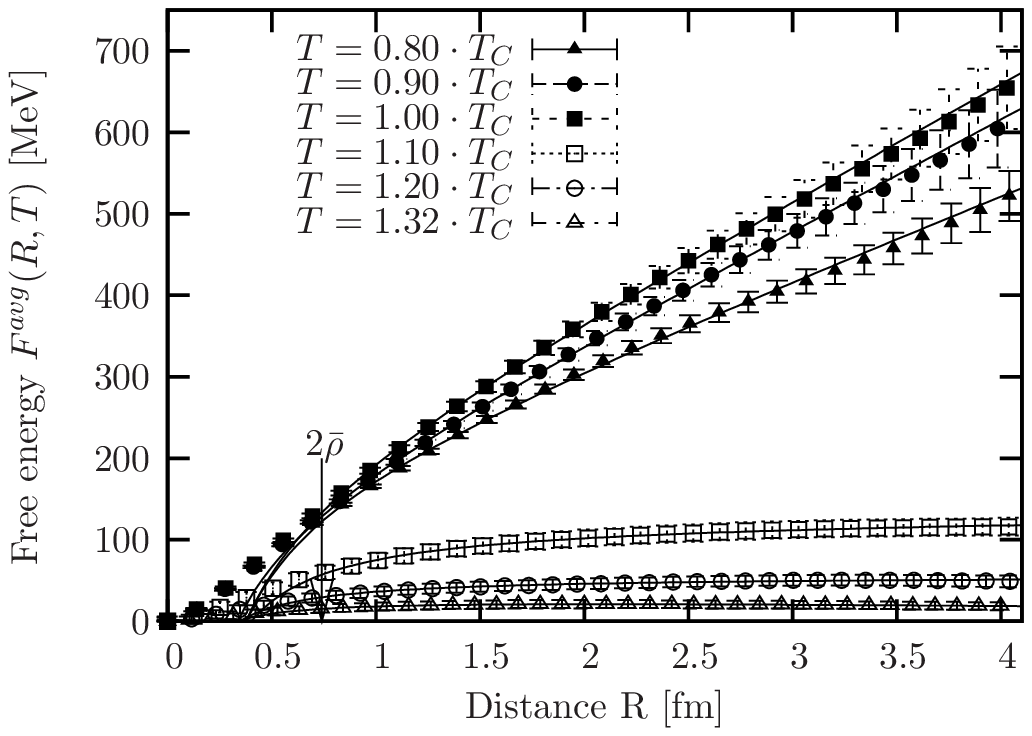}   &   
\includegraphics[angle=0,width=0.48\textwidth]{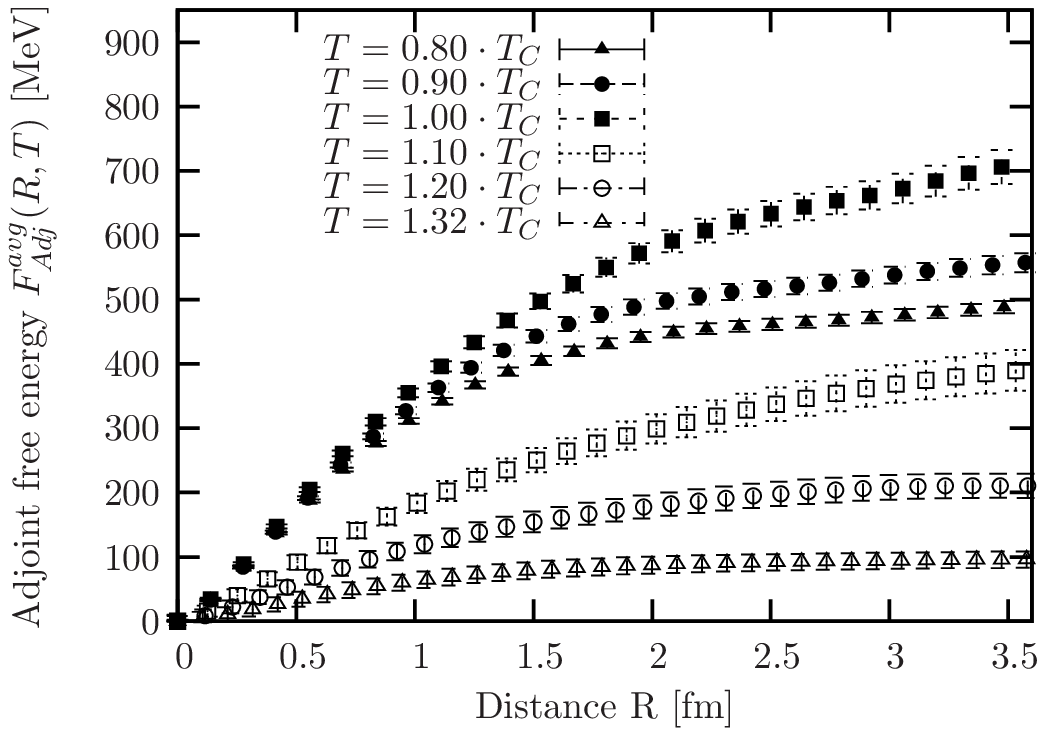}   \\  
\end{tabular}
\caption{color averaged free energy versus distance $R$ at different temperatures 
${T/T_C\fhs{-0.5mm}=\fhs{-0.5mm}0.8,\, 0.9,\, 1.0}$ for the confined and at 
${T/T_C\fhs{-0.5mm}=\fhs{-0.5mm}1.10,\, 1.20,\,1.32}$ for the deconfined phase
in the fundamental~(a) and in the adjoint~(b) representation. 
The fundamental potentials are fitted to (\ref{eq:FitFreeEnergy}) 
for ${R\x\ge\x}2\bar\rho$. This minimal distance is marked.}
\label{fig:60D}
\end{figure*}

With the parameters of the model fixed, we are turning now again to the
question of confinement at finite temperature based on the Polyakov loop.
The color averaged free energy $F^{avg}(R)$ is derived from the Polyakov 
loop correlator by Eq. (\ref{eq:eq2ZZS}) and is presented in 
Fig.~\ref{fig:60D}a. For the temperatures belonging to the confined
phase the short distance behavior of the free energy is independent 
of the temperature. This has also
been observed in lattice studies~\cite{Digal1}. At larger distances 
the curves become linear and dependent
on temperature. What one would expect for the quark-antiquark free energy in the 
confined phase is a linear behavior at large distances. 

This is a consequence of a flux tube forming between the sources. The
flux tube is sometimes understood as an elastic string.
Due to higher string excitations of the color flux tube, one would expect 
corrections to this linear behavior. It remains to be seen whether the
caloron gas model can generate such a behavior besides the linear 
confinement corresponding to a stiff flux tube.

The next-to-leading order correction is a ${-\gamma_L/r}$ term, the 
so-called L\"uscher term~\cite{Luscher1}.
Its coefficient is universally given by ${\gamma_L\x=\x(d-2)\pi/24}$, where
$d$ denotes the space-time dimension. This correction is valid for a large class
of effective bosonic string theories at sufficiently large distances.
We therefore consider the expression
\beq
F^{avg}(R,T) = -\frac{A(T)}{R} + \sigma(T)\cdot R + C(T)
\label{eq:FitFreeEnergy}
\eeq
with the free parameters ${A(T),\,\sigma(T),\,C(T)}$ as the appropriate
fitting function for the measured free energies in the confined phase at not
too small distances.
Our model is based on long-ranged gauge field excitations, and is 
therefore not meant to describe short-ranged physics. As a rough
estimation we assume the model to become valid at distances larger than 
the average caloron diameter, \ie ~for 
${R\fhs{-0.5mm}>\fhs{-0.5mm}2\bar\rho(T)}$. 
We therefore fit the quark-antiquark free energy $F^{avg}(R)$ to 
Eq. (\ref{eq:FitFreeEnergy}) only for these distances, 
as shown in Fig.~\ref{fig:60D}a. One 
sees that the measured free energy agrees very well with 
the expected behavior. 

The string tensions obtained from these fits are listed in Table~\ref{tab:6ZT} 
together with lattice
results~\cite{Digal1} for the quark-antiquark free energy obtained 
for $SU(2)$. The comparison with the lattice 
study shows on one hand that our results for the quark-antiquark free energy are 
of the right order of 
magnitude (at least for ${T/T_C\fhs{-0.5mm}=\fhs{-0.5mm}0.8,\,0.9}$), 
but on the other hand it also reveals that our model is, in its present usage, 
not capable of reproducing the vanishing of the 
time-like string tension at ${T\fhs{-0.5mm}\rightarrow\fhs{-0.5mm} T_C}$. 
A possible cure for this limitation of the model will be discussed 
in the conclusions.
\begin{table*}[htb]
\centering
\begin{tabular}{|c|c|c|c|c|}
\hline
T   &  $\sigma(T)\, [\frac{\mathrm{MeV}}{\mathrm{fm}}]$ & $\frac{\sigma(T)}{\sigma(0)}$ & 
       $\frac{\sigma(T)}{\sigma(0)}$~\cite{Digal1} & $A(T)$  \\ 
\hline
$0.8\, T_C$ & $100.2\pm 2.2$  & $0.31$ & $0.44$ & $0.33\pm 0.02$ \\
$0.9\, T_C$ & $134.1\pm 1.5$  & $0.42$ & $0.29$ & $0.23\pm 0.02$ \\
$1.0\, T_C$ & $139.8\pm 2.1$  & $0.44$ & $0.00$ & $0.32\pm 0.02$ \\
\hline
\end{tabular}
\caption{Below $T_C$: the string tensions $\sigma(T)$ 
and the $1/r$ fit parameters $A(T)$ as obtained from fitting 
the quark-antiquark free energy to (\ref{eq:FitFreeEnergy}).
$\sigma(T)$ is presented in physical units and in units of $\sigma(0)$, 
where $\sigma(0)=318\,$~MeV/fm
for our model according to Section~\ref{subsubsec:FinalFixRho}. 
For comparison, lattice results for $\sigma(T)/\sigma(0)$ taken 
from Ref.~\cite{Digal1} are presented.}
\label{tab:6ZT}
\end{table*}

The corresponding coefficients $A(T)$ belonging to the $1/r$ correction 
term are shown in Table \ref{tab:6ZT} and should be compared with the 
L\"uscher coefficient~\cite{Luscher1} ${\gamma_L\x=\x(d-2)\pi/24} = 0.26$.
It can be seen that the observations are in qualitative agreement with the
size of the L\"uscher term.

While the quark-antiquark free energy rises linearly at large distances 
for the confined phase, the picture changes
completely beyond the phase transition. 
Here, the potentials run into plateaux providing zero string
tension at large distances. This effect is due to the holonomy 
deviating from its maximally non-trivial setting.
In Section~\ref{sec:holonomy_decides} 
we have already seen that the holonomy parameter alone is 
sufficient to determine whether a caloron ensemble
provides a confining potential or not. 
The decreasing caloron density certainly influences the strength 
of the obtained potential but is not responsible for the plateau building
as such. For increasing temperatures the holonomy parameter becomes 
closer to trivial and therefore the plateau building already 
sets in at smaller distances. 

For the deconfined phase we have again fitted the measured potential with 
(\ref{eq:FitFreeEnergy}) just to check the vanishing of the time-like string 
tension. Indeed, almost vanishing string tensions have been obtained.
Additionally, we have also tried to fit the model potentials with some exponentially 
screened expression, since one sees such Yukawa-like behavior in Monte-Carlo
simulations~\cite{Digal1}, but these attempts were unsuccessful.

Additionally, the adjoint charge-anticharge free energy has been determined 
and is shown in Fig.~\ref{fig:60D}b. 
Again, the presented potentials coincide at short distances 
in the confined phase and rise
approximately linearly up to distances 
$R \approx 1\,$~fm. At larger 
distances, the potentials run into plateaux due to string breaking 
as expected. In the 
deconfined phase screening effects are also observed.
\vs{+1mm}
\begin{figure}[bt]
\centering
\includegraphics[angle=0,width=0.48\textwidth]{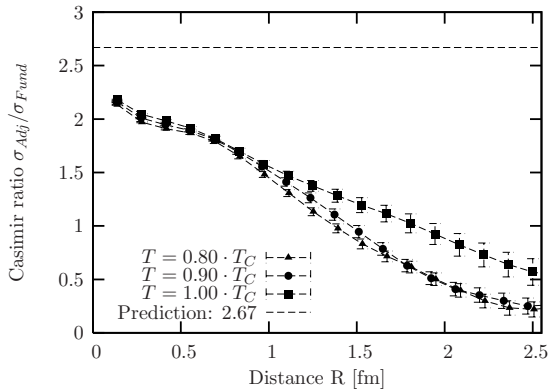}
\caption{Ratio $\sigma_{Adj}/\sigma_{Fund}$ of 
adjoint and fundamental forces from Polyakov loop correlators 
versus the distance $R$
at different temperatures ${T/T_C\fhs{-0.5mm}=\fhs{-0.5mm}0.8,\, 0.9,\, 1.0}$ 
for the confined phase. The theoretical prediction of Casimir scaling is marked 
by the horizontal dashed line.}
\label{fig:60HT}
\end{figure}

Finally, the Casimir scaling hypothesis shall be checked. We show the Casimir ratio 
$\sigma^{adj}(R)/\sigma^{fund}(R)$ together with its 
prediction at short distances in 
Fig.~\ref{fig:60HT}. It can be seen that the measured 
ratio reproduces approximately $85\,\%$
of the theoretical prediction. Here, the deviation from Casimir scaling 
is much stronger 
than in the previously discussed case of the space-like string tension.

\section{Magnetic monopole currents}
\label{sec:monopoles}

In the dual superconductor picture of confinement 
proposed by 't Hooft~\cite{tHooft3} and Mandelstam~\cite{Mandelstam1}, 
the observation of a linearly rising quark-antiquark potential is explained by 
the formation of narrow flux tubes of the chromoelectric field. 
Once such field is inserted to the vacuum by quark sources 
it is squeezed under the influence of a vacuum ``pressure'' provided by a 
Higgs condensate.
Furthermore, 't Hooft also conjectured that the relevant degrees of freedom 
responsible for the confinement property of this theory are actually $U(1)$ 
degrees of freedom. Therefore, one expects the QCD vacuum in the confined phase 
to be populated by random world lines of Abelian magnetic monopoles (Dirac 
monopoles), the origin of the Higgs condensate. If there are electric flux 
tubes, they are enclosed by additional coherent azimuthal magnetic eddy currents.
The necessary condition for the viability of this description is the 
possibility of percolation of Abelian magnetic currents.
It is therefore of interest whether, in the KvBLL caloron gas model, 
percolation of Abelian magnetic currents can be observed.

The Abelian magnetic currents should be observable after fixing an 
Abelian gauge and splitting the gauge field into a (dominating at 
large distances) Abelian component and (important only at short distances) 
non-Abelian components.

The maximally Abelian gauge (MAG) is the most suitable gauge for the purpose
of this decomposition. In this gauge, even the neglect of the non-Abelian 
degrees of freedom is a reasonable approximation. The maximally Abelian 
gauge (MAG) is defined as the gauge maximizing the gauge functional
\beq
R = \sum\limits_{x,\mu} Tr \left[  \tau_3 U_{x,\mu} \tau_3 U^\dagger_{x,\mu}  \right]  
\label{eq:eq7A}
\eeq
by exploiting the non-Abelian gauge freedom. Once the extremization 
is achieved, on average all link variables $U_{x,\mu}$ are as close as 
possible to links
$U^{\prime}_{x,\mu}$ belonging to the Abelian subgroup 
${\{\exp(i\phi\tau_3)\fhs{-0.5mm}:\fhs{-0.5mm}\phi\fhs{-0.5mm}\in\fhs{-0.5mm} 
[0\fhs{-0.5mm}:\fhs{-0.5mm}2\pi]\}\fhs{-0.5mm}\subset \fhs{-0.5mm}SU(2)}$.
The projection 
\beq
U_{x,\mu} = u_0 + i\vec u \vec \tau \rightarrow U^{\prime}_{x,\mu}=
\exp \left(i \theta_{x,\mu} \tau_3 \right)
\label{eq:projection}
\eeq
from $SU(2)$ to $U(1)$ is tantamount to suppressing 
$u_1$ and $u_2$, keeping only the phase 
${\theta_{x,\mu}\fhs{-0.5mm}=
\fhs{-0.5mm}\arg\left(u_0\fhs{-0.5mm} +\fhs{-0.5mm} i u_3\right)}$ 
as the $U(1)$ gauge field.
The $U(1)$ gauge freedom of the remaining Abelian gauge 
field $\theta_{x,\mu}$
remains unfixed. The Abelian monopole currents $j^{mag}_\mu$ 
can be associated with the 
Abelian projected gauge field in the same way as in compact QED~\cite{DeGrand1}. 

Due to magnetic charge conservation 
${\partial_\mu j^{mag}_\mu\fhs{-0.5mm} =\fhs{-0.5mm} 0}$, 
the magnetic currents build closed loops (clusters) on a 
finite and periodic lattice. In Monte Carlo simulations on 
sufficiently large lattices the network of 
monopole clusters was found to be composed of a single very large 
cluster traversing the whole volume and many 
other clusters of small size. 
The histogram $h(l)$ representing the abundance of connected clusters 
containing monopole world lines of total length $l$ (defined 
by counting the occupied links) was found to be split
into two very distinctive parts, $h_{UV}(l)$ for the small clusters 
and $h_{IR}(l)$ for the percolating ones, with a gap separating 
the two~\cite{Hart1,Hart2}.
The percolating cluster alone was shown to reproduce almost the full monopole 
string tension, meaning that the remaining small clusters are not relevant 
for confinement. 
Furthermore, the 3-dimensional (dimensionful) density $n^{(3D)}_{IR}$
of magnetic currents belonging to the percolating cluster has been found 
to be related to the string tension by 
${n^{(3D)}_{IR}\fhs{-0.5mm}/\fhs{-0.5mm}\sigma^{3/2}\fhs{-0.5mm}=\fhs{-0.5mm}0.65(2)}$ 
for MAG~\cite{Bornyakov1}. 
Although the exact ratio depends on the chosen Abelian gauge fixing 
procedure (and is not universal for all possible non-Abelian 
actions)~\cite{Bornyakov2}, 
the percolation itself of magnetic currents, more precisely non-vanishing
winding in spatial directions, is an indication for quark confinement. 

\begin{figure*}[htb]
\centering
\begin{tabular}{cc}
\hs{6mm}(a)& \hs{6mm}(b)\\
\vs{-0mm}\includegraphics[angle=0,height=4.6cm]{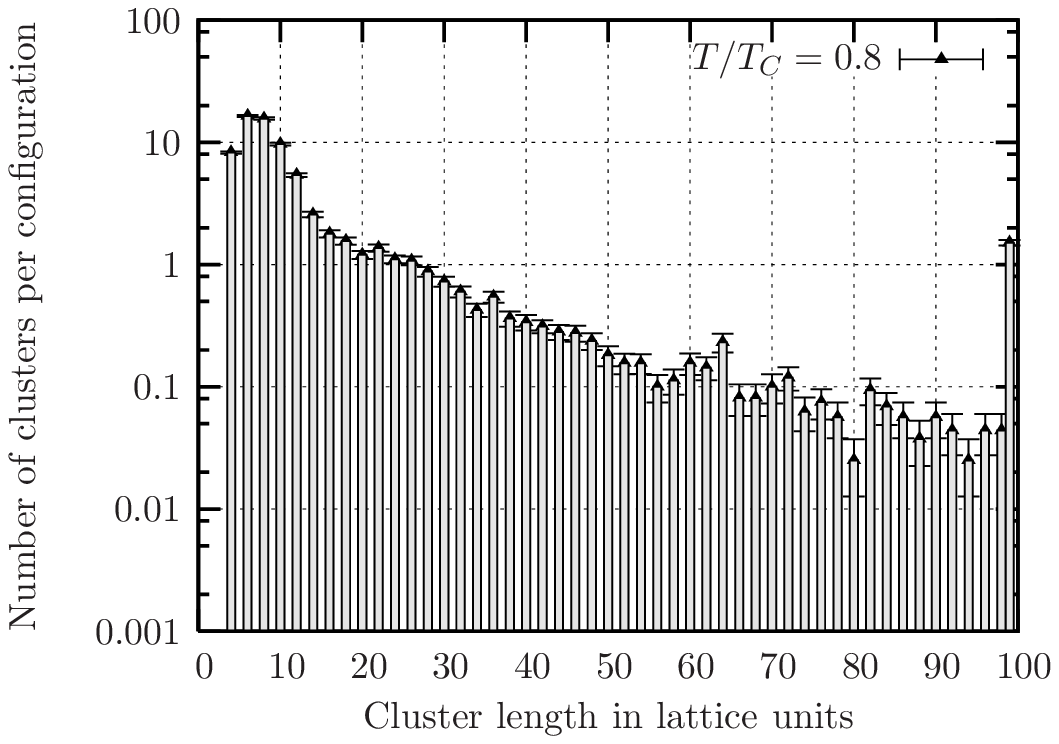} &
\vs{-0mm}\includegraphics[angle=0,height=4.6cm]{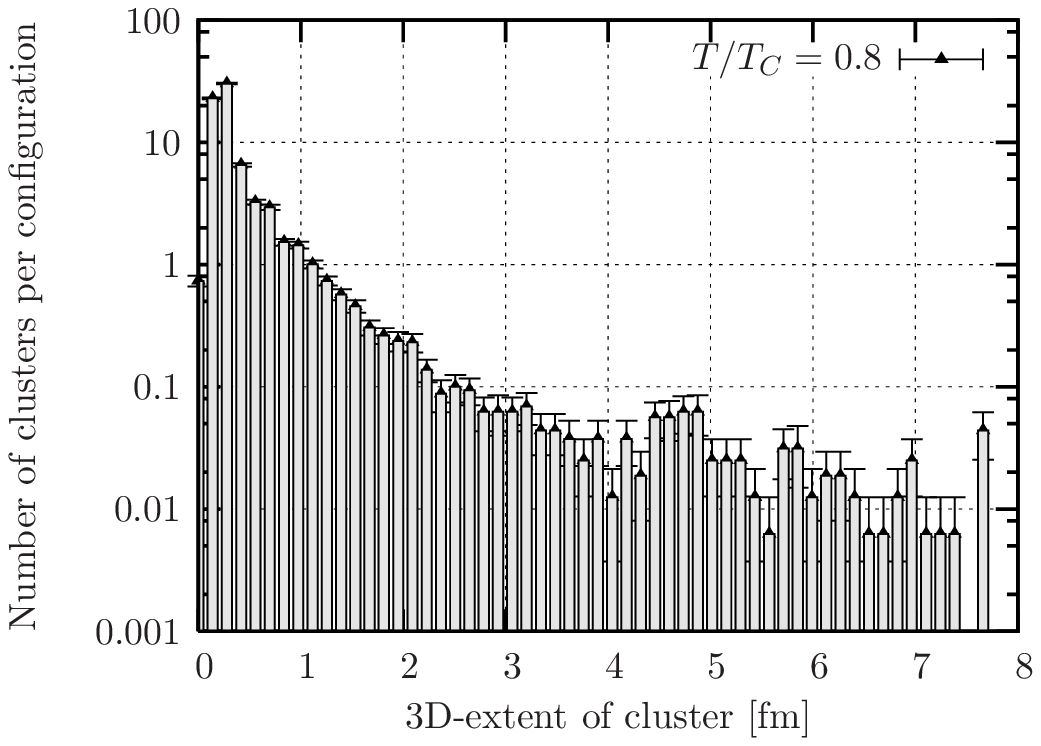}\\
\vs{-0mm}\includegraphics[angle=0,height=4.6cm]{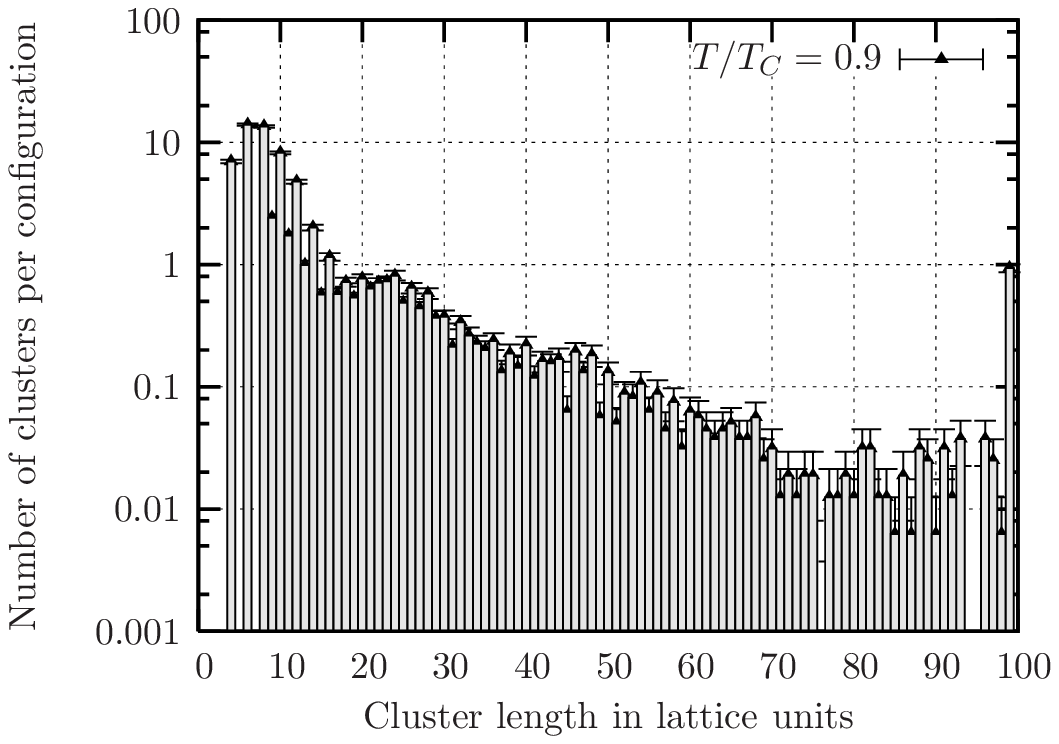} &
\vs{-0mm}\includegraphics[angle=0,height=4.6cm]{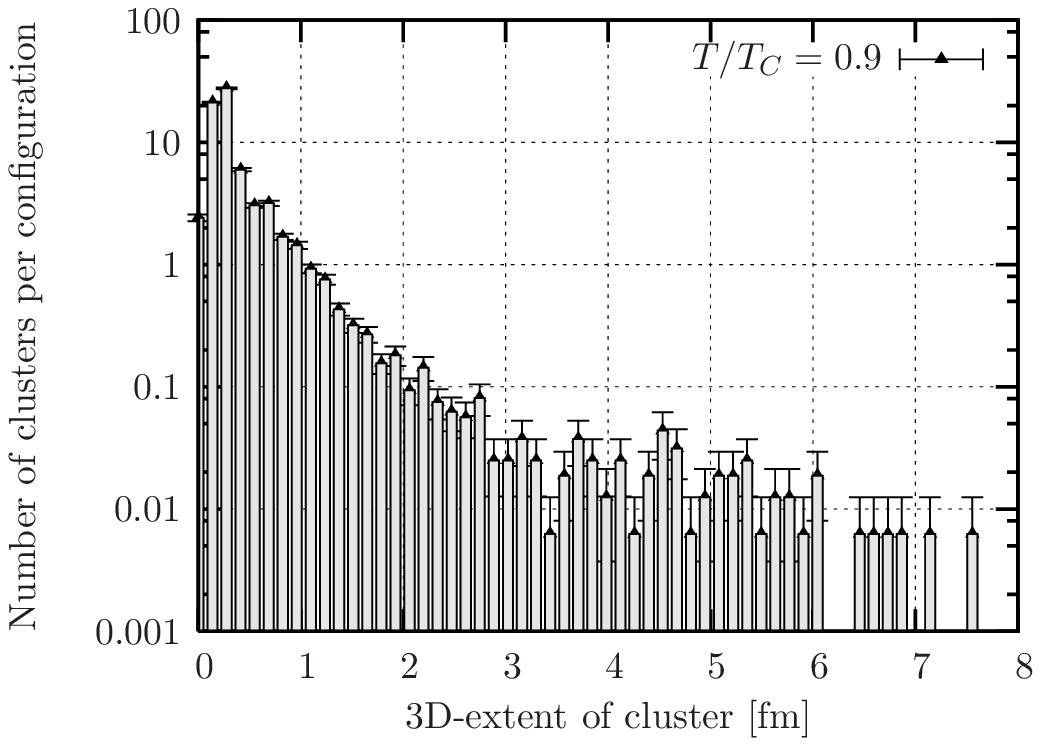}\\
\vs{-0mm}\includegraphics[angle=0,height=4.6cm]{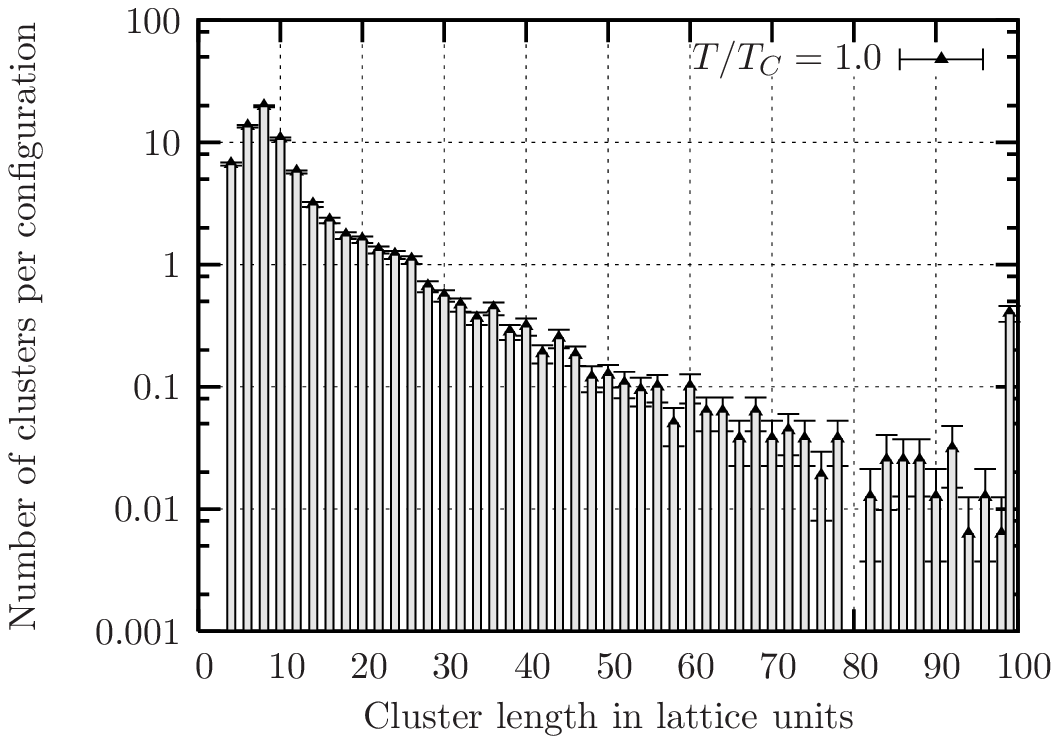} &
\vs{-0mm}\includegraphics[angle=0,height=4.6cm]{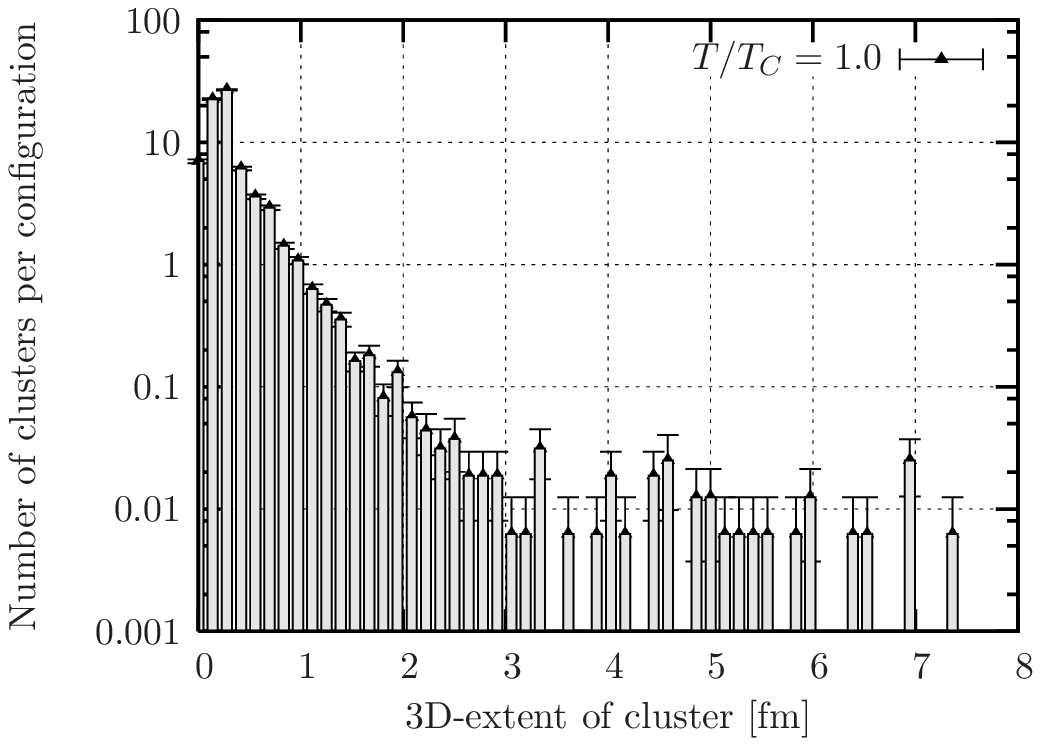}\\
\end{tabular}
\caption{Histograms showing the average number of clusters per configuration 
of a certain cluster size, in (a) characterized by the number of occupied links 
in the cluster, in (b) characterized by the 3D extent of the cluster,
for the temperatures 
${T/T_C\fhs{-0.5mm}=\fhs{-0.5mm}0.8,\, 0.9,\, 1.0}$ in the 
confined phase. Link numbers larger than 100 are set equal to 100.}
\label{fig:7B}
\end{figure*}
Fig.~\ref{fig:7B} shows the average numbers of clusters of a certain size
per configuration as histograms of monopole clusters with respect to their 
length (participating links) in lattice units, $h(l)$, and with respect to 
their maximal 3-dimensional extension, $H(R^{(3D)})$. The histograms have been
evaluated for 160 configurations for each of the temperatures 
${T/T_C\fhs{-0.5mm}=\fhs{-0.5mm}0.8,\, 0.9,\, 1.0}$ in the confined phase.
These measurements have been made on {\it periodic}
lattices, in contrast to the open volume simulations described
above. Since calorons are {\it not} periodic regarding the $\it space-like$
dimensions, special care has to be taken to minimize the resulting action
surplus created by bringing the caloron ensemble onto a periodic lattice.
We therefore compute for every space-time position $x$ the analytic, single 
caloron vector potential ${A_\mu^{(i)}(\Delta x\x\equiv\x x^{(i)}-x)}$, where
$x^{(i)}$ is the position of the $i$-th caloron, guaranteeing by appropriate 
shift operations 
that the inevitable discontinuity of each caloron's vector potential appears
at only those space-time positions (on the antipodal side of the torus) 
where the gauge fields are minimal, thus minimizing the action surplus.
Except for this difference, all model parameters corresponding to these 
selected temperatures are chosen according to Table~\ref{tab:6E} as before.

Due to the periodicity a subtle definition of the cluster extension
$R^{(3D)}$ is needed. At first, the smallest cuboid circumscribing 
the cluster is determined. If the extent of this cuboid is smaller 
than the lattice size in all directions, then the distance between 
all cluster points is unambiguously 
defined. In the opposite case, the cluster can be considered as percolating
(in one or more directions). 
The observed cluster sizes are much larger than the average caloron 
size ${\bar\rho\fhs{-0.5mm}=\fhs{-0.5mm}0.37}$~fm. A small fraction 
of the clusters percolate 
in the sense defined above, and some of these clusters wrap around 
the periodic lattice in the space-like directions, carrying a non-zero
winding number. As discussed above this is a signal for confinement.

In contrast to that, the pictures looks completely different for temperatures 
above $T_c$. 
\begin{figure*}[htb]
\centering
\begin{tabular}{cc}
\hs{6mm}(a)& \hs{6mm}(b)\\
\vs{-0mm}\includegraphics[angle=0,height=4.6cm]{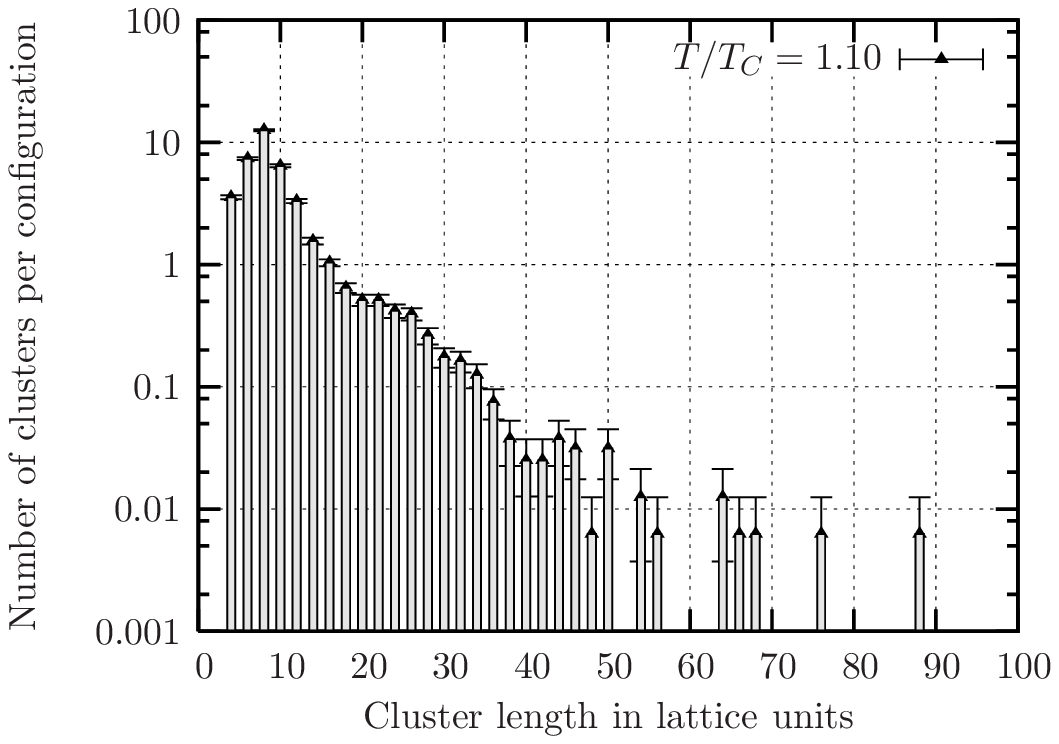} &
\vs{-0mm}\includegraphics[angle=0,height=4.6cm]{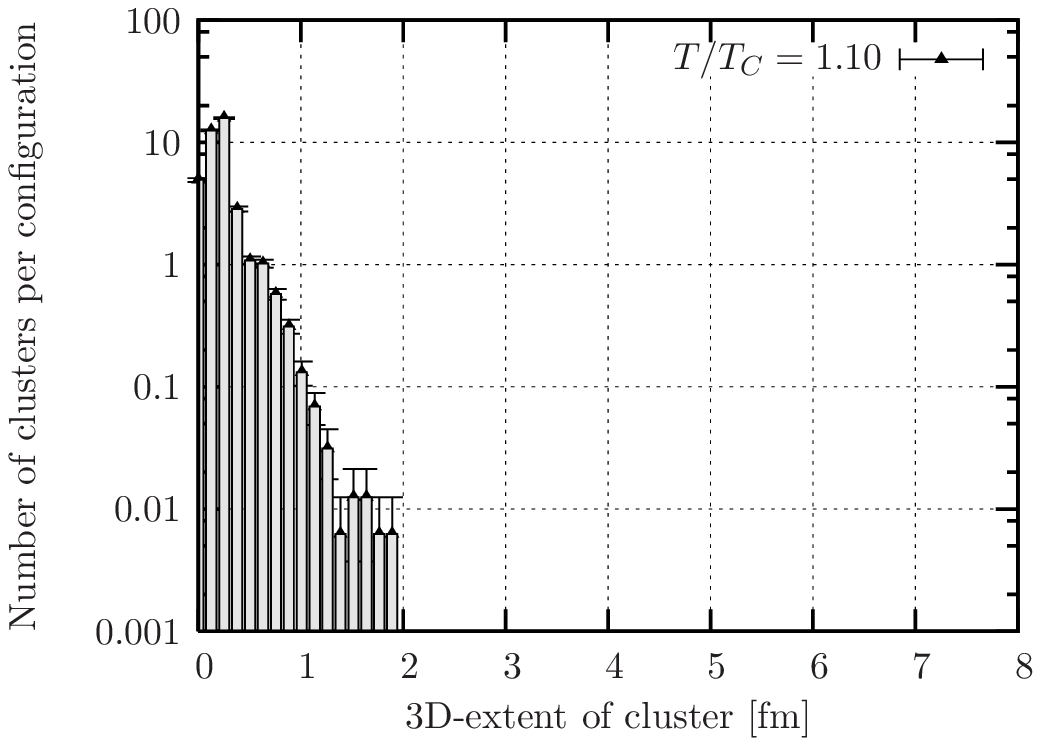}\\
\vs{-0mm}\includegraphics[angle=0,height=4.6cm]{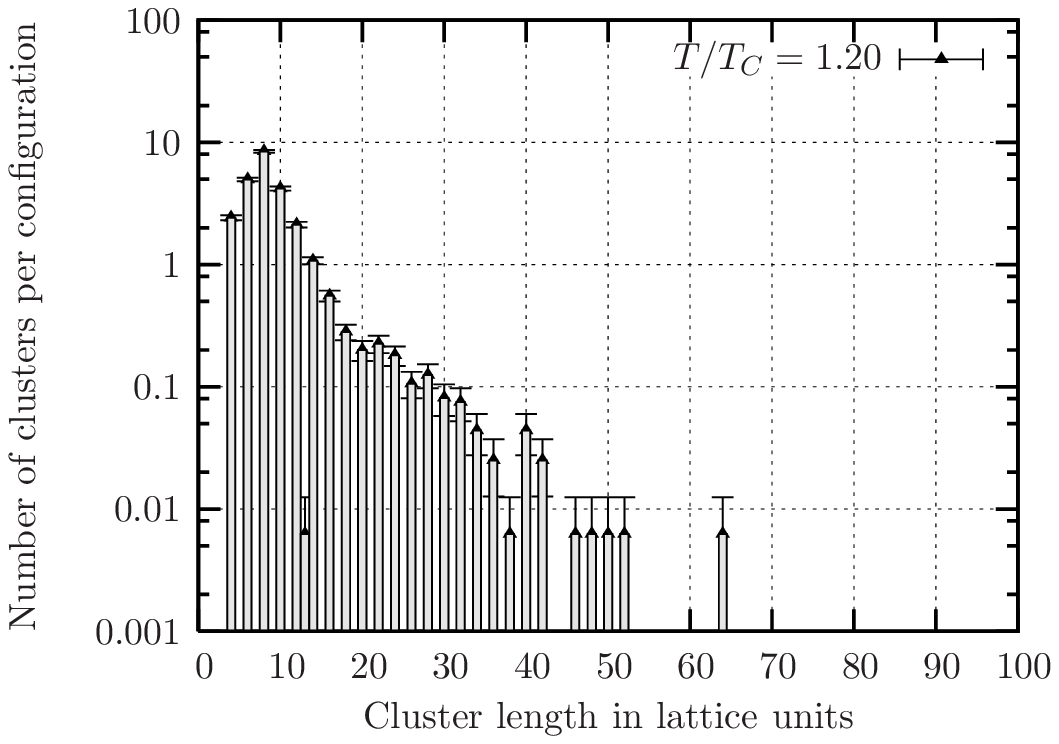} &
\vs{-0mm}\includegraphics[angle=0,height=4.6cm]{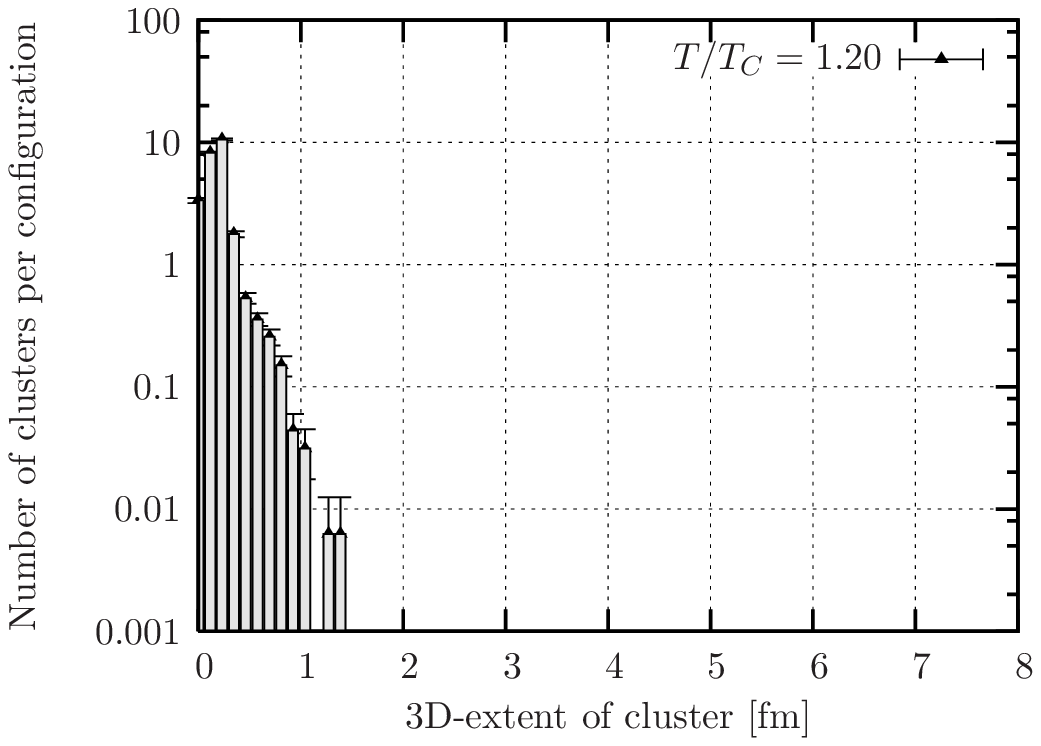}\\
\vs{-0mm}\includegraphics[angle=0,height=4.6cm]{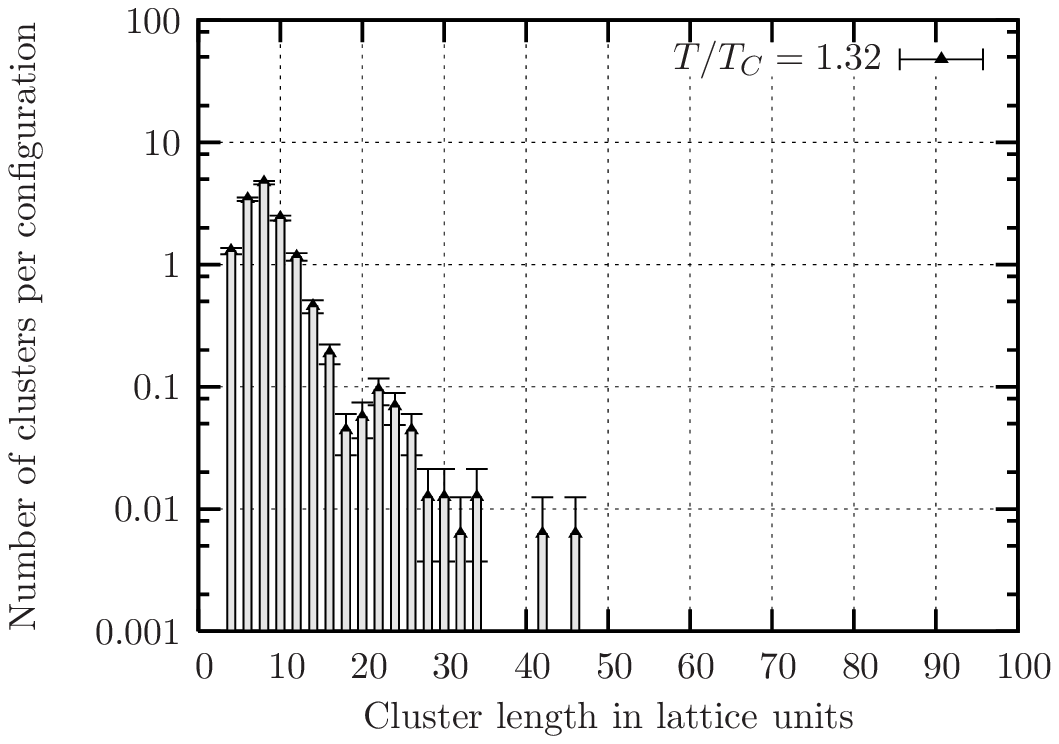} &
\vs{-0mm}\includegraphics[angle=0,height=4.6cm]{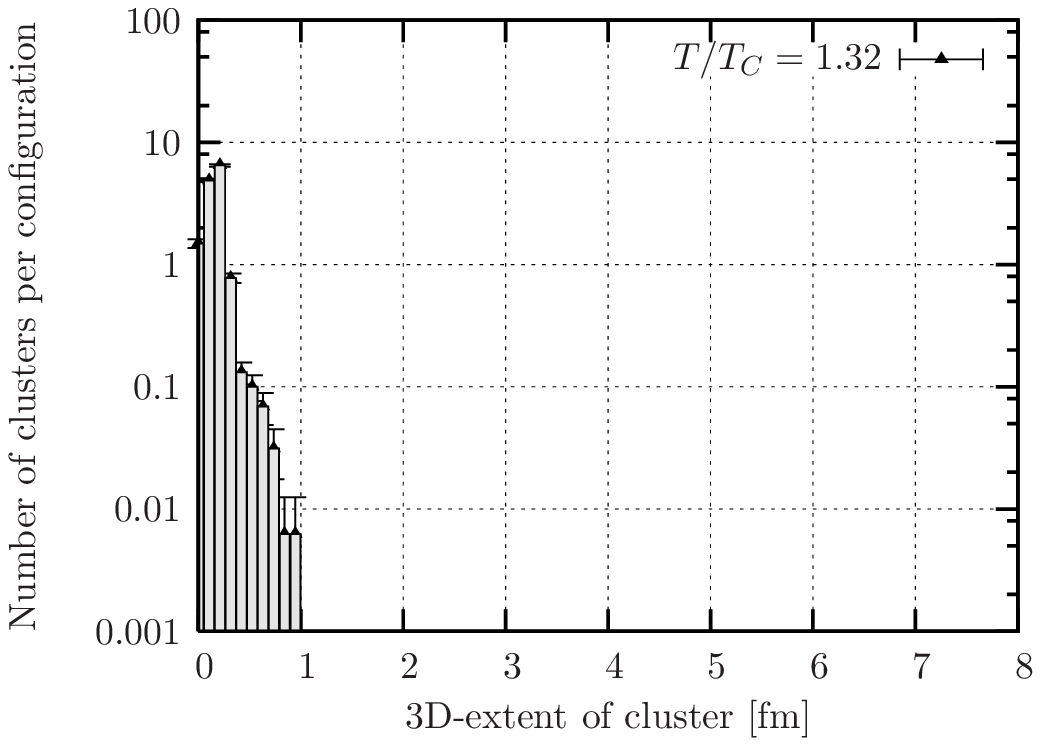}\\
\end{tabular}
\caption{Histograms showing the average number of clusters per configuration 
of a certain cluster size, in (a) characterized by the number of occupied links 
in the cluster, in (b) characterized by the 3D extent of the cluster,
for the temperatures 
${T/T_C\fhs{-0.5mm}=\fhs{-0.5mm}1.10,\, 1.20,\, 1.32}$ in the 
deconfined phase.}
\label{fig:7C}
\end{figure*}
Fig.~\ref{fig:7C} shows the histograms of monopole clusters 
with respect to their length in lattice units, $h(l)$, and with respect to their 
maximal 3-dimensional extension, $H(R^{(3D)})$. Again, 
160 configurations have been evaluated for each of the temperatures 
${T/T_C\fhs{-0.5mm}=\fhs{-0.5mm}1.10,\, 1.20,\, 1.32}$ 
in the deconfined phase. One sees that both distributions are strongly contracted 
towards smaller cluster sizes. No percolation or wrapping around the lattice 
is encountered anymore for temperatures above the critical temperature. 
This observation is in agreement with the breakdown of (time-like) confinement 
that we had found from the study of the Polyakov correlators.

However, it should be noted that the actual sizes $l$ of the percolating 
clusters are one or two orders of magnitude smaller than those seen in 
the Monte Carlo ensembles~\cite{Bornyakov2}. This is only natural since 
the presently evaluated configurations are constructed by superposing single 
calorons. Each non-dissociated caloron gives rise to a closed cluster traversing 
through both of its constituents such that this cluster 
entirely lies within the caloron core. 
For dissociated calorons two disconnected monopole clusters (world lines) 
emerge wrapping around the periodic lattice in time direction. Since the caloron 
gauge field is non-Abelian only in its core, and the vector potential away from 
the core can be approximately identified by the Abelian projection, 
the resulting multi-caloron gauge field is Abelian almost everywhere. 
Since the calorons are sitting relatively close, the monopole world lines 
reorganize themselves into a percolating network without essentially 
increasing the density of monopole currents. Hence, the observed monopole 
cluster sizes in the confinement phase are barely exceeding the minimal cluster 
sizes necessary to facilitate the required monopole percolation. This can be
seen, if one compares the average monopole cluster length in lattice units, as
measured in the simulation, with the average cluster length, that would be obtained,
if each single caloron's monopole cluster was not effected by the presence of 
the other calorons. Table \ref{tab:LinkCountEst} shows simulation results 
together with the estimates for the average cluster length, as obtained from 
the given $\rho$-distribution with $\bar\rho=0.37$~fm. It can be seen that 
the total length of the monopole clusters induced by the single calorons is 
increased only by $25-35\%$ in order to unite closely situated monopole clusters 
to form a connected monopole cluster. 

\begin{table}[htb]
\centering
\begin{tabular}{|c|c|c|}
\hline
 $ T$   &  Measured & Estimated \\ 
\hline
$0.8\, T_C$ &   $14.7$ & $11.5 $ \\
$0.9\, T_C$ &   $14.1$ & $10.9 $  \\
$1.0\, T_C$ &   $12.4$ & $9.2$    \\
\hline
\end{tabular}
\caption{Average cluster lengths in lattice units as measured in the simulations.
The result is compared to an estimate given by the caloron $\rho$ distribution,
assuming each single caloron monopole cluster to be unaltered by the presence of
other calorons.}
\label{tab:LinkCountEst}
\end{table}

\section{Summary and conclusions}
\label{sec:Outlook}

We have introduced a semi-classical model meant to describe the 
Euclidean field histories of (quenched) QCD
as a dilute gas of KvBLL calorons with adjustable holonomy. We have shown, 
that this has the striking impact that the confinement/deconfinement phase 
transition can in principle be modelled by varying nothing but the holonomy
parameter, starting from its maximally non-trivial value 
${\omega(T\fhs{-0.5mm}<\fhs{-0.5mm}T_C)\fhs{-0.5mm} =\fhs{-0.5mm} 0.25}$ 
for the confined phase, towards its trivial setting 
${\omega(T\fhs{-1mm}\gg\fhs{-1mm} T_C)\fhs{-0.5mm}\rightarrow\fhs{-0.5mm} 0}$  or $0.5$, 
respectively, as the temperature increases, such that very high above the 
critical temperature the well-known HS-caloron is finally recovered.

In order to work out the model more quantitatively and in detail,
the parameters of this model have been chosen in agreement with lattice 
observations. Firstly, the holonomy parameter $\omega(T)$ was fixed in 
accordance with lattice studies of the renormalized Polyakov loop in $SU(2)$ 
by exploiting the low-density relation 
${<\fhs{-1mm}|\pol|\fhs{-1mm}> =\fhs{-0.5mm} cos(2\pi |\vec \omega|)}$. 
The reliability of the current determination of $\omega(T)$ could be improved, 
if it would be determined in a self-consistent manner. However, as long as the 
holonomy is maximally non-trivial in the confined phase and becomes trivial 
only at high temperatures above $T_C$, the exact functional form of $\omega(T)$ 
has only little influence on the observed ``asymptotic'' string tensions.

Second, the caloron density $n(T)$ has been determined by lattice results 
for the topological susceptibility $\chi(T)$ by assuming
calorons and anticalorons to be uncorrelated. Since we do not take any 
caloron/anticaloron correlation into account for the sampling of their 
positions, this is the adequate assumption with respect to our model, 
although calorons and anticalorons are actually known to be correlated. 
This has lead us to adopt a constant caloron density 
${n(T)\fhs{-0.5mm}=\fhs{-0.5mm}1}$~fm$^{-4}$ 
as usual in the confined phase and a decreasing density 
${\propto\fhs{-0.5mm}\chi(T)}$ in the deconfined phase. 
This is not compelling since the decreasing topological susceptibility could 
also be explained by topologically uncharged objects becoming dominant, 
such as caloron-anticaloron molecules, although there is no independent
indication for that to happen in quenched QCD. 

Also in another respect the current model is incomplete in the deconfined 
phase because it does not consider non-perturbative configurations 
of other origin. This incompleteness becomes obvious by 
seeing the space-like string tension decreasing above the critical 
temperature due to the reduced caloron density. This is unphysical 
since the space-like string tension is known to rise at higher temperatures,
completely independent from the topological susceptibility which becomes
tiny.
Apart from the loophole forcing us to accept a decreasing density of calorons,
it could well be that monopoles being neither self-dual nor anti-self-dual 
build the spatial string tension throughout the deconfined phase.

The breakdown of the physically confining potentials above the 
critical temperature in our model is
driven by the holonomy alone. The decrease
of the caloron density would not be required.

Third, the $\rho$-distribution was determined by reinterpreting the single 
caloron quantum weight in the dense-packing limit. Due to its divergence at 
${\rho\fhs{-0.5mm}\rightarrow\fhs{-0.5mm}\infty}$ for non-trivial holonomy, 
additional arguments rendering the $\rho$-distribution finite had to be 
invoked. Since its analytical form could only be qualitatively motivated, 
a more sound investigation of the $\rho$-distribution in a dense medium of
calorons would be worthwhile. However, having fixed the analytical 
form of the size distribution, the average caloron size in the confined phase
has been unambiguously determined by comparing the space-like string tensions 
with those obtained in lattice studies. Hence, the usual criticism of 
instanton/caloron based models, namely that almost every desired result could be 
obtained from this kind of models by suitably setting the average caloron size, 
is actually not applicable here. 

The color-averaged quark-antiquark free energy was then extracted from the 
Polyakov loop correlation in multi-caloron configurations constructed according 
to a simple superposition scheme. Since our model is based on long-ranged gauge 
field excitations of average diameter ${2\bar\rho(T)}$, the model is not supposed 
to yield reasonable results at distances smaller than 
${R\fhs{-0.5mm}<\fhs{-0.5mm}2\bar\rho(T)}$. 
For ${R\fhs{-0.5mm}>\fhs{-0.5mm}2\bar\rho(T)}$ the obtained free 
energies show the expected linear plus L\"uscher term behavior up to distances of 
${R\fhs{-0.5mm}\approx\fhs{-0.5mm} 4}$~fm in the confined phase. Above the critical
temperature the obtained potentials run into plateaux corresponding to the deconfinement
of quarks. 
The obtained string tensions are of the correct order of magnitude as can be 
seen by a comparison with lattice studies. However, our model is, so far, 
not capable to  reproduce the correct temperature dependence of the time-like 
(color-averaged) string tension, which is known to gradually decrease with rising
temperature before it is going to vanish 
at ${T\fhs{-0.5mm}=\fhs{-0.5mm} T_C}$. This limitation is caused by 
the fact that our model does not allow for mixed holonomies within a single gauge 
field configuration. A minimal improvement of the model would be to sample the 
uniform caloron holonomies in the confined phase according to a distribution with 
non-vanishing, temperature dependent width (in accordance to what is seen 
in lattice Monte Carlo simulations) rather than being fixed exactly to zero 
($\omega=0.25$). 

By studying the potentials in the adjoint representation screening effects 
due to string breaking could be observed. For small distances 
a ratio between the fundamental and adjoint effective string tensions 
of ${\sigma^{adj}/\sigma^{fund}\fhs{-0.5mm} \approx\fhs{-0.5mm} 2.2}$ 
was obtained, which is close to but 
smaller than the ratio $8/3$ predicted by the Casimir scaling hypothesis. 
For space-like string tensions screening effects could also be seen and 
the Casimir hypothesis was even better fulfilled. 
Here, a ratio ${\sigma^{adj}/\sigma^{fund}\fhs{-0.5mm} \approx\fhs{-0.5mm} 2.6}$ 
could be observed for all selected confining temperatures. 

Finally, the formation of large magnetic monopole clusters was studied in the 
caloron gas model. Percolating clusters traversing the whole lattice volume 
were observed for the confining phase, which is a necessary condition 
for confinement. For the deconfined phase the distribution of monopole cluster 
sizes is contracted towards smaller cluster sizes and no spatially percolating 
clusters are found anymore, which is in agreement with Monte Carlo lattice 
observations. 
Not unexpectedly, the observed monopole clusters are one or two orders of 
magnitude smaller (in path length) than those obtained in Monte Carlo studies.
The average cluster lengths in the caloron ensembles are only slightly increased 
by rearrangements induced by the caloron superposition compared to the 
sum of the cluster sizes of the separate calorons.

In conclusion we would like to say, despite the limitations that our model 
still suffers from, the KvBLL caloron gas model offers an interesting scenario for 
the confinement/deconfinement phase 
transition. With the proposed improvements, the KvBLL caloron gas model most 
probably will give a consistent picture of the phase transition. 
Therefore it is a worthwhile topic to further investigation.

\vs{-4mm}
\section*{Acknowledgements}

We would like to thank Boris Martemyanov for introducing the idea how to superpose
KvBLL calorons with non-trivial holonomy. We are grateful to Falk Bruckmann for 
sharing his insights on KvBLL caloron systems and for enlightening discussions and 
correspondence. We appreciate the support from Pierre van Baal, having
organized a meeting in Leiden one year ago, and his recent helpful comments on 
this paper prior to publication. We thank Olaf Kaczmarek for drawing 
our attention to Ref.~\cite{Digal1} and the authors of this work for the kind 
permission to use their results in Fig.~\ref{fig:5E}. Finally, we are grateful 
to Alexander Veselov for providing the code for MAG fixing based on the simulated 
annealing algorithm. 

\bibliographystyle{unsrt}
\bibliography{bibliography}

\end{document}